\documentclass[
reprint,
longbibliography,
amsmath, 
amssymb, 
aps, 
superscriptaddress,
pra
]
{revtex4-2}

\usepackage{graphicx}% Include files
\usepackage{dcolumn}% Align table columns on decimal point
\usepackage{bm}% bold math
\usepackage{braket} %For the representation of bra and ket
\usepackage{hyperref}% add hypertext capabilities
\usepackage{subfigure} % subgraficos
\usepackage{nccmath}
\usepackage{amsthm} % Theorem Formatting
\usepackage{mathrsfs} %for the Hilbert symbol
\usepackage{xfrac} %for the fraction 1/2
\usepackage{pifont}
\usepackage[export]{adjustbox}
\usepackage{bbm} %probability font
\usepackage[utf8]{inputenc} 
\usepackage[normalem]{ulem}
\usepackage[usenames,dvipsnames,svgnames,table]{xcolor}
\usepackage[sort&compress]{natbib}
\usepackage{color}
\usepackage{longtable}
 
\usepackage{commath}
\usepackage{multirow}

\begin{document}
\title{Method for Generating Randomly Perturbed Density Operators Subject to Different Sets of Constraints}

\author{J. A. Monta\~nez-Barrera}
 	\altaffiliation{ja.montanezbarrera@ugto.mx (J.A. Monta\~nez-Barrera)}
	\affiliation{Department of Mechanical Engineering, Universidad de Guanajuato, Salamanca, GTO 36885, Mexico}
\author{R. T. Holladay}
	\affiliation{Department of Mechanical Engineering, Virginia Tech, Blacksburg, VA 24061, USA}	
\author{G. P. Beretta}
	\affiliation{Mechanical and Industrial Engineering Department, Universit$\grave{a}$ di Brescia, via Branze 38, 25123 Brescia, Italy}	
	
\author{Michael R. von Spakovsky}
	\altaffiliation{vonspako@vt.edu (M.R. von Spakovsky)}
	\affiliation{Department of Mechanical Engineering, Virginia Tech, Blacksburg, VA 24061, USA}

\begin{abstract}
This paper presents a general method for producing randomly perturbed density operators subject to different sets of constraints. The perturbed density operators are a specified ``distance" away from the state described by the original density operator. This approach is applied to a bipartite system of qubits and used to examine the sensitivity of  various entanglement measures on the perturbation magnitude. The constraint sets used include constant energy, constant entropy, and both constant energy and entropy. The method is then applied to produce perturbed random quantum states that correspond with those obtained experimentally for Bell states on the IBM quantum device ibmq\_manila. The results show that the methodology can be used to simulate the outcome of real quantum devices where noise, which is important both in theory and simulation, is present.

\begin{description}
	\vspace{0.2cm}
	\item[Keywords] IBMQ; Qiskit; quantum computation; entanglement; random quantum states; perturbed quantum states
\end{description}

\end{abstract}

\maketitle

\section{\label{int}Introduction}

Quantum information and computing systems rely on the phenomena of superposition and entanglement to efficiently perform certain tasks that are otherwise inefficient when performed on classical computers. While any quantum system may exhibit superposition, only composite quantum systems (i.e., those consisting of two or more subsystems) may exhibit entanglement.

For a composite of two subsystems only the correlated states that are not separable are entangled. Mathematically, a density operator $\hat{\rho}_{AB}$ represents correlated states of the two subsystems $A$ and $B$ if

\begin{equation}
	\hat{\rho}_{AB} \neq \hat{\rho}_A\otimes\hat{\rho}_B
\end{equation}
where $\hat{\rho}_{A(B)} = \mathrm{Tr}_{B(A)}(\hat{\rho}_{AB})$ while it represents  separable states if there exist positive-valued probabilities, $w_i$,  and   pure density operators for the subsystems, $\hat{\rho}_{i}^{A}$ and $\hat{\rho}_{i}^{B}$, such that

\begin{equation}
	\hat{\rho}_{AB} = \sum_i w_i\, \hat{\rho}_{i}^A \otimes \hat{\rho}_{i}^B
\end{equation}
While there is no fundamental definition quantifying how entangled two systems are, numerous metrics exist that aim to capture system entanglement \cite{Horodecki2009}. Examples include the mutual information, concurrence, and the Clauser-Horne-Shimony-Holt (CHSH) operator expectation value,  each of which provide useful insight into the states of composite quantum systems. 

Because each of these metrics quantify entanglement in a different way, they exhibit different behavior relative to the physical property changes of the composite system. Thus, one of the focuses of this paper is to gain insight into the effect of the physical state of the system on various entanglement measures and, in particular, on the sensitivity of these measures to perturbations of the state $\hat{\rho}_{AB}$ of the composite system.

Generating random quantum states (RQS) is an important tool for understanding quantum systems with many degrees of freedom, and it has a variety of uses in quantum information science. For example and principally, it allows one to understand, characterize, and parametrize quantum states \cite{Grondalski2002, Radtke2008, Bruning2012,Maziero2015}. In this work, we extend RQS generation to perturbed states via the development and implementation of a tool for characterizing and simulating perturbed Bell and Bell-diagonal states that correspond to the behavior of state preparation in Noisy Intermediate-Scale Quantum (NISQ) devices \cite{Preskill2018}. The NISQ devices are characterized for having different error sources grouped into two main categories: i) decoherent errors such as relaxation or dephasing (see, e.g., \cite{Lang2010, Cole2010, Mazzola2010, Maziero2010, Xu2010, Pramanik2019, Montanez-Barrera2020}) and ii) coherent errors such as gate, crosstalk, and readout errors \cite{Krinner2020, Figueroa-Romero2021, Parrado-Rodriguez2021}. Our approach can be used for error simulation in quantum devices, which is important for methods of error mitigation where various techniques are used to reduce errors in noisy models \cite{OBrien2021, Strikis2021}. Usually, these errors of depolarization, amplitude damping, or phase damping are simulated using perturbation shifting in the unitary matrices that describe a process on a quantum device. Another possible application is for initial state preparation for simulations of a quantum device using a master equation \cite{Montanez-Barrera2020, Riera-Campeny2021}. In Section \ref{application}, an application of our methodology is presented in which the methodology is able to obtain the same range of Bell states prepared on IBM's superconducting quantum processor, $ibmq\_manila$.

The methodology proposed here is used to randomly perturb an arbitrary baseline density operator $\hat{\rho}_0$ subject to an arbitrary set of constraints on the expectation values associated with the density operator. This method is initially illustrated by perturbing a bipartite system where each subsystem has only two levels but may be straightforwardly extended to composite systems of arbitrary numbers of subsystems. To understand the effect of perturbations on entanglement, the baseline density operator, $\hat{\rho}_0$, which represents the state of the composite system consisting of subsystems $A$ and $B$, can be chosen, for example, to be a Bell diagonal state represented by \cite{Liu2016}

\begin{equation}\label{Belldiagonal1}
	\hat{\rho}_0 = \frac{1}{4} \sum_{i = 0}^3 c_i\,  \hat{\sigma}_{i}^A\otimes\hat{\sigma}_{i}^B
\end{equation}
where $\hat{\sigma}_0 = \hat{I}$ is the identity operator and  $\hat{\sigma}_1 = \hat{\sigma}_X$, $\hat{\sigma}_2 = \hat{\sigma}_Y$, and $\hat{\sigma}_3 = \hat{\sigma}_Z$ are the three Pauli spin operators and for this work, the scalar coefficient $c_0$ must be $c_0 = 1$, to represent a Bell diagonal state, and the other coefficients are chosen as $c_1 = 0.996$, $c_2 = 0.4$, and $c_3 = -0.4$. A Bell diagonal state is chosen for this initial illustration because it is a state of non-zero entropy that can be written directly as a sum of Bell states (which are maximally entangled pure states of the composite system). Thus, after the general density operator perturbation methodology and the setup of the experiment are presented in Section \ref{Sec::Chap_4_Ptb_Model}, perturbations of the baseline state $\hat{\rho}_0$ are generated under four different sets of constraints. Properties of the resulting perturbed states are presented in Section \ref{Sec::Chap_4_Results} as well as simulations of the Bell states of the $ibmq\_manila$ quantum processor. Finally, the main trends exhibited in these sets of perturbations are discussed in Section \ref{Sec::Chap_4_Conclusion} along with possible follow-up work.

\section{Methodology}
\label{Sec::Chap_4_Ptb_Model}

\subsection{General Perturbation Approach and ``Closeness" of Quantum States} \label{Sec::generalperturbation}

This section presents a method for generating a randomly perturbed density operator whose distance from the original density operator is O$\left(\epsilon\right)$ where the notion of distance between two states is discussed in Section \ref{SSec::Chap_4_Dist_Entang_Meas} and $\epsilon$ is a specified parameter. This approach permits the enforcement of arbitrary sets of constraints on the expectation values of the resulting perturbed operator, while ensuring that the entire space of states in the neighborhood of the original state is uniformly sampled. 

While this approach can be applied to composite systems consisting of any numbers of subsystems, it is illustrated here using a bipartite system of two-level subsystems. To begin with, recall that given any density operator there is a unique  positive semidefinite $\sqrt{\hat{\rho}}$ satisfying $\rho=\sqrt{\hat{\rho}}\sqrt{\hat{\rho}}$, called the non-negative square root of $\hat{\rho}$. Also recall that an arbitrary operator $\hat{\gamma}$ with Hermitian conjugate $\hat{\gamma}^\dagger$ points to the two density operators  $\hat{\gamma}\hat{\gamma}^\dagger/\mathrm{Tr}(\hat{\gamma}\hat{\gamma}^\dagger)$ and $\hat{\gamma}^\dagger\hat{\gamma}/\mathrm{Tr}(\hat{\gamma}^\dagger\hat{\gamma})$. The random perturbation strategy proposed here starts by computing the non-negative square root  of the baseline density operator $\hat{\rho}$ to be perturbed,
\begin{equation}
	\hat{\gamma}_0 = \sqrt{\hat{\rho}_0}
\end{equation}
Then a set of operators $\hat{\gamma}_r$ are randomly generated that belong to a neighborhood of $\hat{\gamma}_0$ in the space of linear operators on the Hilbert space of the composite system subject to the desired constraints and the normalization constraint $\mathrm{Tr}(\hat{\gamma}_r^\dagger\hat{\gamma}_r)=1$. As a result, the operators $\hat{\rho}_r=\hat{\gamma_r}\hat{\gamma_r}^\dagger$ are automatically positive semi-definite (non-negative eigenvalues) by construction and are taken as the desired set of the perturbed density operator. 

To setup the perturbation procedure, consider a composite of $M$ qubits, so that the Hilbert spaces associated with the subsystems are all two-dimensional. The four-dimensional space of linear operators on each of these Hilbert spaces is spanned by the Hermitian basis $\{\hat\sigma_0/\sqrt{2},\hat\sigma_1/\sqrt{2},\hat\sigma_2/\sqrt{2},\hat\sigma_3/\sqrt{2}\}$ --- orthonormal with respect to the symmetrized Hilbert-Schmidt scalar product $\hat{\alpha}\cdot\hat{\beta}=\mbox{\textonehalf}\mathrm{Tr}(\hat{\alpha}^\dagger\hat{\beta} + \hat{\beta}^\dagger\hat{\alpha})$ ---  Therefore,  the linear combinations 
\begin{equation}
\frac{1}{2^{M/2}}	\sum_{i_1,...,i_{M}= 0}^{3} \eta_{i_1,..,i_{M}}\bigotimes_{K = 1}^M\hat{\sigma}_{i_K}
\end{equation}
represent the non-Hermitian  operators on the Hilbert space of the composite system if at least one of the coefficients $\eta_{i_1,...,i_M}$ is a complex number and the Hermitian operators if they are all real numbers. Here, $\eta_{i_1,..,i_{M}}$ is the random perturbation for the combination of operators $\hat{\sigma}_{i_1},..,\hat{\sigma}_{i_M}$ where $i_n \in \{0,1,2,3\}$. Note that the $\frac{1}{2^{M/2}}$ pre-factor inserted here and in similar equations  below is necessary to make the basis $\frac{1}{2^{M/2}}	\bigotimes_{K = 1}^M\hat{\sigma}_{i_K}$  orthonormal. 

Now, assuming that the composite system is bipartite with subsystems $A$ and $B$, the above orthonormal basis for the real space of Hermitian operators on the Hilbert space of the composite system is $\frac{1}{2}\hat{\sigma}_{i}^{A} \otimes\hat{\sigma}_{j}^{B}$ and, therefore, any Hermitian operator $\hat{\omega}$  can be written as
\begin{equation}\label{genericoperator}
	\hat{\omega}=\frac{1}{2} \sum_{i,j = 0}^3\eta[\hat{\omega}]_{ij}\,\hat{\sigma}_{i}^A\otimes\hat{\sigma}_{j}^B
\end{equation}
and represented by the 4$\times$4 matrix of coefficients $\eta[\hat{\omega}]_{ij}$, which in turn can be obtained by direct projection onto the operator basis elements, 
\begin{equation}
\eta[\hat{\omega}]_{ij} =\frac{1}{2} \mathrm{Tr}[\hat{\omega}\,(\hat{\sigma}_{i}^{A} \otimes\hat{\sigma}_{j}^{B})]
\end{equation}
For example, Bell diagonal states, which may also be conveniently expressed as 
\begin{equation}\label{Belldiagonal2}
	\hat{\rho}_0 =\frac{1}{2}\left[\begin{array}{cccc}
		a^2+b^2 & 0 & 0 & b^2-a^2 \\
		0 & c^2+d^2 & c^2-d^2 & 0 \\
		0 & c^2-d^2 & c^2+d^2 & 0 \\
		b^2-a^2 & 0 & 0 & a^2+b^2
	\end{array}\right]
\end{equation}
are represented by the coefficients
\begin{equation}\label{Belldiagonalcoeff}
	\eta[\hat{\rho}_0]_{ij} =  \frac{1}{2}\delta_{ij}c_i=\frac{1}{2}\left[\begin{array}{cccc}
		c_0 & 0 & 0 & 0 \\
		0 & c_1 & 0 & 0 \\
		0 & 0 & c_2 & 0 \\
		0 & 0 & 0 & c_3
	\end{array}\right]
\end{equation}
and their non-negative square roots
\begin{equation}
	\hat{\gamma}_0 =\frac{1}{2}\left[\begin{array}{cccc}
		a+b & 0 & 0 & b-a \\
		0 & c+d & c-d & 0 \\
		0 & c-d & c+d & 0 \\
		b-a & 0 & 0 & a+b
	\end{array}\right]
\end{equation}
by the coefficients
\begin{equation}\label{eta0}
	\eta[\hat{\gamma}_0]_{ij} =\frac{1}{2}\left[\begin{array}{cccc}
		a\mbox{+}b\mbox{+}c\mbox{+}d &\hskip-14pt 0 &\hskip-14pt 0 &\hskip-14pt 0 \\
		0 &\hskip-14pt b\mbox{$-$}a\mbox{+}c\mbox{$-$}d &\hskip-14pt 0 &\hskip-14pt 0 \\
		0 &\hskip-14pt 0 &\hskip-14pt a\mbox{$-$}b\mbox{+}c\mbox{$-$}d &\hskip-14pt 0 \\
		0 &\hskip-14pt 0 &\hskip-14pt 0 &\hskip-14pt a\mbox{+}b\mbox{$-$}c\mbox{$-$}d
	\end{array}\right]
\end{equation}
where $a$, $b$, $c$, $d$ are the eigenvalues of $\hat{\gamma}_0$ and in terms of the parameters $c_0$, $c_1$, $c_2$, $c_3$ used in Eqs. (\ref{Belldiagonal1}) and (\ref{Belldiagonalcoeff}) are given by
\begin{equation}
\begin{aligned}
	a=\mbox{\textonehalf}\sqrt{c_0\mbox{$-$}c_1\mbox{+}c_2\mbox{+}c_3}\qquad b=\mbox{\textonehalf}\sqrt{c_0\mbox{+}c_1\mbox{$-$}c_2\mbox{+}c_3}\\
	c=\mbox{\textonehalf}\sqrt{c_0\mbox{+}c_1\mbox{+}c_2\mbox{$-$}c_3}\qquad d=\mbox{\textonehalf}\sqrt{c_0\mbox{$-$}c_1\mbox{$-$}c_2\mbox{$-$}c_3}
\end{aligned}
\end{equation}
and, of course, $\mathrm{Tr}(\hat{\rho}_0)=a^2\mbox{+}b^2\mbox{+}c^2\mbox{+}d^2=1$.

The above representation of Bell diagonal states makes it easy also to generate them with given eigenvalues or randomly. For the latter purpose it suffices to pick four independent numbers  $a_r$, $b_r$, $c_r$, $d_r$ in the range -1 to 1, and then normalize them by $R=\sqrt{ a_r^2\mbox{+}b_r^2\mbox{+}c_r^2\mbox{+}d_r^2}$ so as to obtain the parameters $a=a_r/R$, $b=b_r/R$, $c=c_r/R$, $d=d_r/R$ for use with Eq. (\ref{Belldiagonal2}), while for use in Eq. (\ref{Belldiagonal1}) the parameters are
\begin{equation}
	\begin{aligned}
		c_0&=a^2\mbox{+}b^2\mbox{+}c^2\mbox{$+$}d^2\\ 
		c_1&=\mbox{$-$}a^2\mbox{+}b^2\mbox{+}c^2\mbox{$-$}d^2\\ 
			c_2&=a^2\mbox{$-$}b^2\mbox{+}c^2\mbox{$-$}d^2\\
			c_3&=a^2\mbox{+}b^2\mbox{$-$}c^2\mbox{$-$}d^2
	\end{aligned}
\end{equation}

For our procedure, Eq. (\ref{genericoperator}) is used  to express the operators $	\hat{\gamma}_\epsilon - \hat{\gamma}_{0}$ where each $\hat{\gamma}_\epsilon$ is in the Hermitian neighborhood of the baseline $\hat{\gamma}_0$ and is selected by randomly picking the values of the sixteen real coefficients $ \eta_{ij}$ from a suitable distribution. 
The Hilbert–Schmidt norm $\epsilon=\sqrt{\mathrm{Tr}[(\hat{\gamma}_\epsilon - \hat{\gamma}_{0})^2]}=(\sum_{i,j = 0}^3\eta_{ij}^2)^{1/2}$ can be taken as a measure of the distance from the baseline operator and provides guidance on the random choices made so as to populate and/or be confined to the neighborhood as desired. For example, when the $\eta_{ij}$'s are all taken from a zero-mean normal distribution $\mathcal{N}(0, \tilde\sigma)$ with standard deviation $\tilde\sigma$, the value of $\epsilon/\tilde\sigma$ is distributed according to the well-known $\chi$ distribution.

The perturbed operators 
\begin{equation}\label{gammaepsilon}
	\hat{\gamma}_\epsilon = \hat{\gamma}_{0} + \frac{1}{2} \sum_{i,j = 0}^3\eta_{ij}\,\hat{\sigma}_{i}^A\otimes\hat{\sigma}_{j}^B
\end{equation}
are Hermitian but no longer the square root of true density operators because $\mathrm{Tr}(\hat{\gamma}_\epsilon^2) \neq 1$. Moreover, no constraints have yet been enforced on the expectation values of the resulting perturbed density operators. 

To impose the unit-trace condition and the other constraints, each $\hat{\gamma}_\epsilon$ is corrected by subtracting from it the symmetrized gradient of each constraint computed at $\hat{\gamma}_\epsilon$ multiplied by an undetermined multiplier. To fix ideas, assume that constant energy and constant entropy constraints are to be imposed. Then, for each randomly generated $\hat{\gamma}_\epsilon$, the Hermitian operator
\begin{equation}
	\begin{aligned}
	\hat{\gamma}_r &= \hat{\gamma}_\epsilon - \{\hat{I},\hat{\gamma}_\epsilon\}\lambda_I -  \{\hat{H},\hat{\gamma}_\epsilon\}\lambda_H -
 \{-\hat{I}-\hat{B}\ln\hat{\gamma}^2_\epsilon,\hat{\gamma}_\epsilon\}	\lambda_S\\
	&=(1-2\lambda_I+2\lambda_S)\hat{\gamma}_\epsilon - \{\hat{H},\hat{\gamma}_\epsilon\}\lambda_H-
	 \{-\hat{B}\ln\hat{\gamma}^2_\epsilon,\hat{\gamma}_\epsilon\}\lambda_S
	\label{Eq::Chap_4_gam_r_gam_eps_ex}
\end{aligned}
\end{equation}
is defined and points to the desired constrained perturbed density operator  $\hat{\rho}_r=\hat{\gamma}^2_r$ which is obtained once the multipliers $\lambda_I$, $\lambda_H$, and $\lambda_S$ are determined by imposing the respective constraints: C1 = $\mathrm{Tr}(\hat{\gamma}^2_r) = 1$, C2 = $\mathrm{Tr}(\hat{\gamma}^2_rH) = \mathrm{Tr}(\hat{\gamma}^2_0H)$, and C3 = $\mathrm{Tr}(\hat{\gamma}^2_r\ln\hat{\gamma}^2_r) = \mathrm{Tr}(\hat{\gamma}^2_0\ln\hat{\gamma}^2_0)$. In Eq. (\ref{Eq::Chap_4_gam_r_gam_eps_ex}), $\{\cdot,\cdot\}$ denotes the usual anticommutator, while operator $\hat{B}$, following \cite{Beretta2009}, is the projector onto the range of $\hat{\gamma}^2_\epsilon$ (i.e., onto the subspace spanned by the eigenvectors of $\hat{\gamma}^2_\epsilon$ with nonzero eigenvalues).

In general, it may be of interest to impose $N$ (possibly nonlinear) constraints of the form $\mathrm{Tr}[\hat{\gamma}^2\hat{C_i}(\hat{\gamma}^2)]=\mathrm{Tr}[\hat{\rho}_0\hat{C_i}(\hat{\rho}_0)]$, where the always necessary unit-trace constraint is included by setting $\hat{C_1}=\hat{I}$.   The symmetrized gradients have the form $\{\hat{G_i}(\hat{\gamma}^2),\hat{\gamma}\}$ where $\hat{G_i}(\hat{\gamma}^2)=\hat{C_i}(\hat{\gamma}^2)+\mbox{\textonehalf}\{\hat{\gamma}^2,\hat{C'_i}(\hat{\gamma}^2)\}$. For example, as seen above, for the unit-trace constraint: $\hat{C_1}=\hat{G_1}=\hat{I}$; for the energy constraint: $\hat{C_2}=\hat{G_2}=\hat{H}$; and for the entropy constraint: $\hat{C_3}(\hat{\gamma}^2)= -\hat{B}\ln\hat{\gamma}^2$ ,  $\hat{C'_3}(\hat{\gamma}^2)= -\hat{B}\hat{\gamma}^{-2}$, and  $\hat{G_3}(\hat{\gamma}^2)=(-\hat{I}-\hat{B}\ln\hat{\gamma}^2$). Therefore,  the non-negative square root of the desired perturbed density operator, $\hat{\gamma}_r$, can be written as
\begin{equation}
	\hat{\gamma}_r = \hat{\gamma}_\epsilon - \sum_{i = 1}^N  \{\hat{G}_i(\hat{\gamma}^2_\epsilon),\hat{\gamma}_\epsilon\}\,\lambda_i
	\label{Eq::Chap_4_gam_r_gam_eps}
\end{equation}
where the $N$ multipliers $\lambda_i$ are to be determined by substituting Eq. \ref{Eq::Chap_4_gam_r_gam_eps} into the  system of constraint equations
\begin{equation}
	 \mathrm{Tr}\left(\hat{\gamma}_r^2\,\hat{C}_i(\hat{\gamma}_r^2)\right)=\mathrm{Tr}\left(\hat{\rho}_0\,\hat{C}_i(\hat{\rho}_0)\right) 
	\label{Eq::Chap_4_gam_r_gam_eps_cons}
\end{equation}
and solving numerically. An illustration of the procedure for determining the multipliers for a single constraint in addition
to the necessary unit-trace constraint is presented in Sec. \ref{example1} below.

 Note that while for some sets of constraints it may be possible to find an analytical solution for $\hat{\gamma}_r$, this is not possible for the constant entropy and other  constraints that are nonlinear in the state operator.  However, the resulting system of $N$ Eqs. (\ref{Eq::Chap_4_gam_r_gam_eps_cons}) can be solved numerically for the values of the $\lambda_i$ where it is noted that solutions near  $\lambda_i = 0$ are sought.

After the set of constraints is applied, the new set of $\eta^{Cn}_{ij}$ can be recovered by

\begin{equation}
	\eta^{Cn}_{ij} = \eta[\hat\gamma_r^{Cn} - \hat\gamma_0]_{ij} = \frac{1}{2} \mathrm{Tr} \left((\hat{\gamma}_r^{Cn}-\hat\gamma_0)\,(\hat{\sigma}_{i}^{A} \otimes\hat{\sigma}_{j}^{B})\right)
	\label{Eq18}
\end{equation} 
where $Cn$ is the constraint set applied (i.e., $C_1$ or $C_1$,$C_2$ or $C_1$,$C_3$ or $C_1$,$C_2$,$C_3$) and $\eta^{Cn}_{ij}$ is computed for $i,j \in \{0,1,2,3\}$ so that the $\hat{\gamma}_r^{Cn}$ can be expressed in the form
\begin{equation}
\hat{\gamma}_r^{Cn} = \hat{\gamma}_{0} + \frac{1}{2} \sum_{i,j = 0}^3\eta^{Cn}_{ij}\hat{\sigma}_{i}^{A}\otimes\hat{\sigma}_{j}^{B}
\end{equation}
Eq. (\ref{Eq18}) is a simple projection onto the basis elements that allows the computation of the coefficients directly. An illustration of the procedure for determining the explicit solution of the $\eta[\hat\gamma_r]_{i,j}$ for the case of no non-trivial constraints is given in Sec. \ref{Constraintschapunittrace} below.

\subsection{Review of State Distance and Entanglement Measures}
\label{SSec::Chap_4_Dist_Entang_Meas}

To understand the effects of the perturbations, measures characterizing the ``closeness" of the perturbed state to the baseline state are first examined. This is followed by measures quantifying the entanglement of the two subsystems. 

To begin with, a widely used measure of the closeness of two quantum states $\hat{\rho}_1$ and $\hat{\rho}_2$ is the (square-root) fidelity defined according to \cite{Nielsen_Chuang2010} as the trace norm of the product of the respective non-negative square roots $\hat{\gamma}_1=\sqrt{\hat{\rho}_1}$ and $\hat{\gamma}_2=\sqrt{\hat{\rho}_2}$ such that
\begin{equation}
	F(\hat{\rho}_1,\hat{\rho}_2) = \|\hat{\gamma}_1\hat{\gamma}_2\|_1= \mathrm{Tr}\left(\abs{\hat{\gamma}_1\hat{\gamma}_2}\right)=\mathrm{Tr}\sqrt{\hat{\gamma}_1\hat{\gamma}_2^2\hat{\gamma}_1}
\end{equation}
It is noted that 	$0\le F(\hat{\rho}_1,\hat{\rho}_2) = 	F(\hat{\rho}_2,\hat{\rho}_1) \le 1 $ and that $F(\hat{\rho}_1,\hat{\rho}_2) = 1$ if and only if the two states are identical. It is also noteworthy that some authors call fidelity the square of $F$.

Alternative measures of fidelity can be obtained by using two other related measures of  distance between quantum states based, following \cite{Beretta2009}, on the observation that  density operators map one-to-one to the unit sphere, $\|\hat{\gamma}\|=1$, in the real space of Hermitian operators on the Hilbert space of the system equipped with the real scalar product $\hat{\gamma}_1\cdot\hat{\gamma}_2=\mathrm{Tr}(\hat{\gamma}_1\hat{\gamma}_2)$ and the norm $\|\hat{\gamma}\|=\mathrm{Tr}(\hat{\gamma}^2)$. The distance between two points on the surface of a sphere can be equivalently measured by the geodesic arc length $d$, the chord length $c$, and the central angle $\theta$. For the unit sphere, $d=\theta$, $c=2\sin(\theta/2)$ and, in terms of the unit-norm 'vectors' $ \hat{\gamma}_1$ and $\hat{\gamma}_2$ associated with the two points, $\cos(\theta)=\hat{\gamma}_1\cdot\hat{\gamma}_2$ and $c^2=\|\hat{\gamma}_1-\hat{\gamma}_2 \|^2=2(1-\hat{\gamma}_1\cdot\hat{\gamma}_2)$. Therefore,  the following  expressions (and related identities) provide geometrically well-founded measures of distance between two state operators:
\begin{equation}
	\begin{aligned}
	\theta(\hat{\rho}_1,\hat{\rho}_2) &= \arccos(\hat{\gamma}_1\cdot\hat{\gamma}_2)
	=\arccos\left(\mathrm{Tr}(\hat{\gamma}_1\hat{\gamma}_2))\right)\\
	&= 2\arccos\left(1-
	\frac{1}{2}\mathrm{Tr}\left((\hat{\gamma}_1-\hat{\gamma}_2)^2\right)\right)\\
	&= 2\arcsin\left(
	\frac{1}{\sqrt{2}}\sqrt{1-\mathrm{Tr}(\hat{\gamma}_1\hat{\gamma}_2)}\right)\\
		&= 2\arcsin\left(
	\frac{1}{2}\sqrt{\mathrm{Tr}\left((\hat{\gamma}_1-\hat{\gamma}_2)^2\right)}\right)\\
	c(\hat{\rho}_1,\hat{\rho}_2) &=\sqrt{ \mathrm{Tr}\left((\hat{\gamma}_1-\hat{\gamma}_2)^2\right)} = \sqrt{2 \left(1-\mathrm{Tr}(\hat{\gamma}_1\hat{\gamma}_2)\right)}\\
	\end{aligned}
\end{equation}
where $-1\le\mathrm{Tr}(\hat{\gamma}_1\hat{\gamma}_2) \le 1$, $0\le \theta\le\pi$, $0\le c\le 2$, $\theta$ and $c$ are zero if and only if the two states are identical, and $\theta\approx c$ for small values. When operator $\hat{\gamma}_1-\hat{\gamma}_2$ is expressed in the form of Eq. (\ref{genericoperator}) then $c$ is the Hilbert–Schmidt norm $(\sum_{i,j = 0}^3\eta_{ij}^2)^{1/2}$ and $\theta=2\arcsin(c/2)$, and the coefficients can be obtained by direct projection onto the operator basis elements, 
$\eta_{ij} =\frac{1}{2} \mathrm{Tr} [(\hat{\gamma}_1-\hat\gamma_2)\,(\hat{\sigma}_{i}^{A} \otimes\hat{\sigma}_{j}^{B})]$. Although beyond the scope of the present paper, we mention that the distance measures $\theta$ and $c$ may be easily extended to more complex composite systems by means of the multipole approach to quantum state representation developed in the almost forgotten pioneering papers \cite{Band1970, Park1971, Band1971}.

To examine the effects of the proposed perturbation strategy on the correlations and  entanglement of the resulting perturbed state, three quantifiers are considered: the mutual information, the concurrence, and the CHSH operator maximum expectation value. The mutual information, $I(\hat{\rho})$, which is related to the entropy of the system, is computed as \cite{Nielsen_Chuang2010}
\begin{equation}
	I(\hat{\rho}) = -\mathrm{Tr}_A(\hat{\rho}_A\ln(\hat{\rho}_A)) - \mathrm{Tr}_B(\hat{\rho}_B\ln(\hat{\rho}_B)) + \mathrm{Tr}(\hat{\rho}\ln(\hat{\rho}))
\end{equation}
It is  a non-negative quantity, equal to zero only when the states of subsystems $A$ and $B$ are uncorrelated and hence separable, i.e., when $\hat{\rho}=\hat{\rho}_A\otimes\hat{\rho}_B$. However, it is noted that it being nonzero is not necessarily indicative of entanglement.

The concurrence, $C(\hat{\rho})$, as given by \cite{Hill1997,Wooters1998}, is an entanglement monotone (meaning that it increases as entanglement increases) and for two-qubit states is expressed as 
\begin{equation}
	C(\hat{\rho}) = \max(0,r_1 - r_2 - r_3 - r_4)
\end{equation}
where the $r_i$'s are the eigenvalues,  in decreasing order, of the operator $\hat{R}$ defined as
\begin{equation}
	\hat{R}(\hat{\rho}) = \sqrt{\hat{\gamma}\,\tilde{\rho}(\hat{\rho})\,\hat{\gamma}}
\end{equation}
where $\hat{\gamma}=\sqrt{\hat{\rho}}$ is the non-negative square root of $\hat{\rho}$ and $\tilde{\rho}(\hat{\rho})$ is obtained by first computing the  complex conjugate $\hat{\rho}^*$ of  $\hat{\rho}$ in the standard basis  and then spin-flipping it with respect to the Pauli matrix $\hat{\sigma}_1$ expressed in the same basis, i.e.,
\begin{equation}
	\tilde{\rho}(\hat{\rho}) = \left(\hat{\sigma}_1\otimes\hat{\sigma}_1\right)
	\hat{\rho}^*
	\left(\hat{\sigma}_1\otimes\hat{\sigma}_1\right)
\end{equation}

The final entanglement measure considered here is the violation of the CHSH inequality \cite{Clauser1969}. This inequality, which is closely related to Bell's inequality \cite{Bell1964}, states that local hidden variable theories cannot predict correlations above a value of 2, and, thus, correlations above this value must be due to the quantum phenomenon of entanglement. Cirel'son \cite{Cirelson1980} showed that when accounting for quantum mechanical effects, the maximum value of the correlations exhibited between the subsystems in a composite quantum system is $2\sqrt{2}$. In an experiment, it is possible to compute various expectation values for the CHSH operator, $\hat{B}_{\rm CHSH}$, depending on the relative orientation of the experimental measurement. However, an analytical form for the maximum possible expectation value of this operator for a given state of the system can be computed from \cite{Horodecki2009}
\begin{equation}
\langle\hat{B}_{\rm CHSH}(\hat{\rho})\rangle_{max} = 2\sqrt{t_{11}^2 + t_{22}^2}
\end{equation}
where $t_{11}$ and $t_{22}$ are the two largest eigenvalues of the 3$\times$3 matrix $T_{\hat{\rho}}^T T_{\hat{\rho}}$, with the elements of $T_{\hat{\rho}}$  given by
$t_{ij} = \mathrm{Tr} \left(\hat{\rho}\,(\hat{\sigma}_i \otimes\hat{\sigma}_j)\right)$ for $i,j \in \{1,2,3\}$.

Before presenting results for these distance and entanglement measures relative to the proposed perturbation methodology, the next section provides a description of the IBM transmon quantum device $ibmq\_manila$ used here to experimentally obtain approximate Bell states.

\subsection{IBM Transmon Quantum Device}
\label{setup}
The $ibmq\_manila$ device used in the experiments is a superconductive quantum processor with a Falcon architecture, 5 qubits, a quantum volume (QV) of 32, an average T$_1$=160.55 $\mu s$, an average T$_2$=59.29 $\mu s$, and an average CNOT error of 9.730e-3. Qubits q0 and q1 are employed with $f_0 = 4.963$ GHz and  $f_1 = 4.838$ GHz. The Hamiltonian of the two-qubit system is expressed as  
\begin{equation}
\hat H = -\frac{\hbar}{2}\left(\omega_0\, \hat\sigma_3 \otimes \hat\sigma_0 + \omega_1\, \hat\sigma_0 \otimes \hat\sigma_3 \right)
\end{equation}
where $\omega_0=2\pi f_0$ and $\omega_1=2\pi f_1$.

\begin{figure}[!htp]
	\hspace*{-0.6cm}\includegraphics[width=6cm]{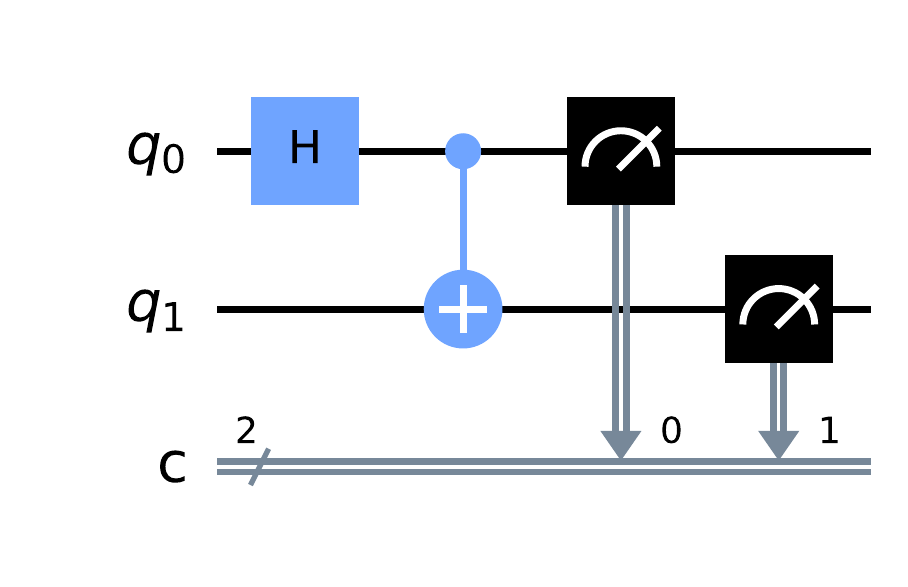}\\
	\centering a)\\
	\hspace*{-0.6cm}\includegraphics[width=10cm, height=8cm]{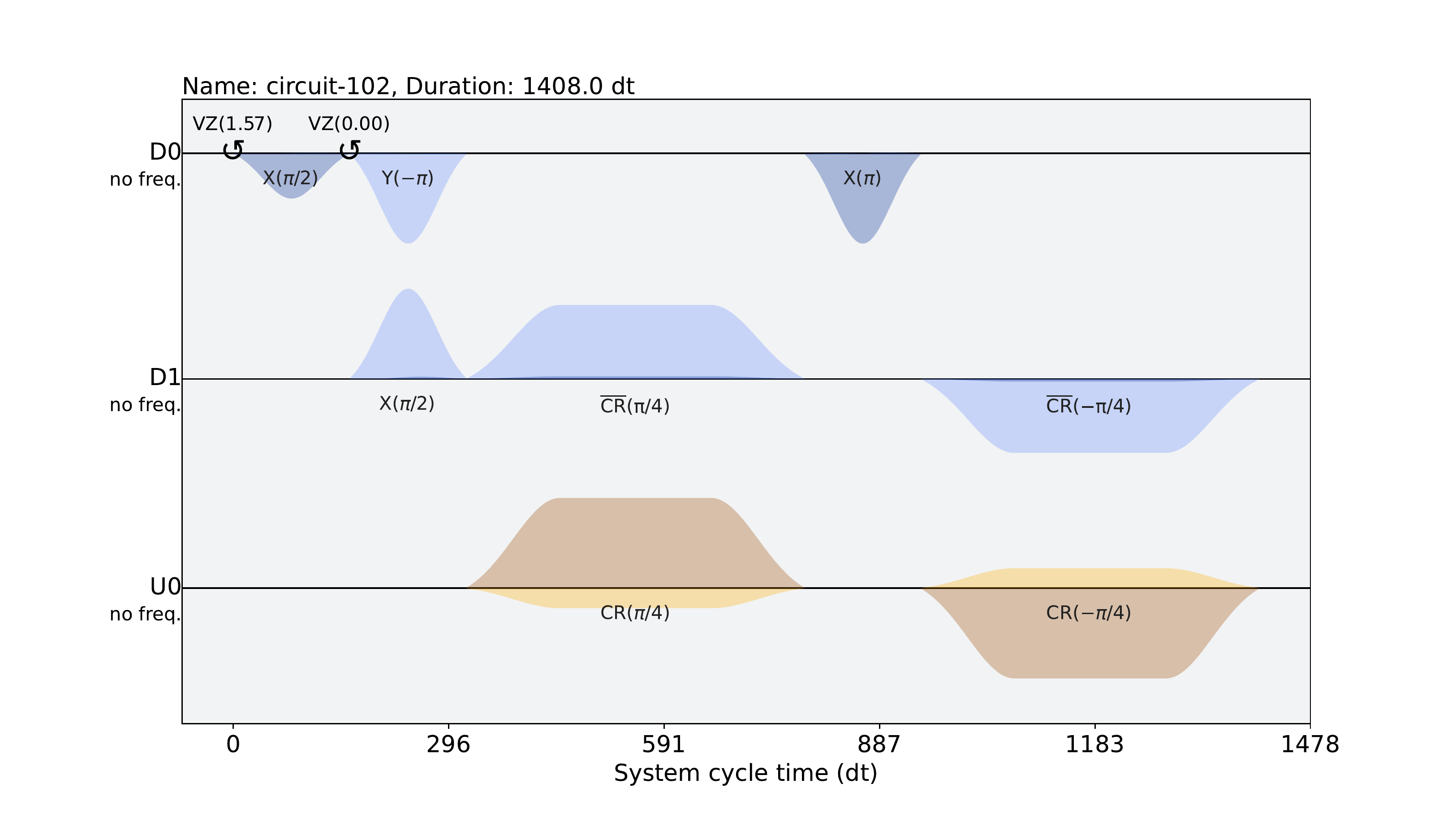}\\
	\centering b)
	\caption{\label{BellCirPulses}a) Circuit to obtain a Bell state is comprised of a Hadamard gate in the control qubit followed by a CNOT gate; b) set of pulses for the IBM transmon quantum device $ibmq\_manila$ that represent the circuit in a).}
\end{figure}

The experimental setup in $ibmq\_manila$ is the sequence of gates presented in Fig. \ref{BellCirPulses} a). Ideally, such gates result in the Bell state $\left | \Phi^+ \right> = \frac{1}{\sqrt{2} }\left(\left | 00 \right>  + \left | 11 \right> \right)$. The set of gates (a Hadamard and a CNOT gate) shown in this figure are translated into pulse-level control as shown in Fig. \ref{BellCirPulses} b) where D0 and D1 are the driven channels for qubit 0 and qubit 1, respectively. The U0 channel permits the qubits to interact with a cross resonance interaction and is the principal element for the CNOT gate.

To reconstruct the density operator $\hat\rho_R^n$ of the Bell state preparation, the procedure of Smolin $et\;al.$ \cite{Smolin2012} is followed where the density operator $\hat\rho_R^n$ is recovered with 2000 shots or measurements in each of 9 different Pauli bases. A shot is defined as a single repetition of a given circuit.  Next, for each experimental result, the non-negative square root operator, $\hat{\gamma}_R^n=\sqrt{\hat\rho_R^n}$, is computed first after which the non-negative square root  $\hat\gamma_0 = \sqrt{\hat\rho_0}$ of the ideal Bell state 
\begin{equation}
	\hat{\rho}_0 = \left | \Phi^+ \right> \left < \Phi^+ \right| 
	\label{rho0}
\end{equation}
is subtracted. 

Two hundred approximate Bell states via 3,600,000 measurements (2000x9x200) are generated using the transmon quantum device $ibmq\_manila$. A set of random values for the $\eta_{i,j}$ are determined using a normal distribution $\mathcal{N}(\mu_{i,j}, \sigma_{i,j})$. This set of values is then used in
\begin{equation}
\hat{\gamma}_\epsilon = \hat{\gamma}_{0} + \frac{1}{2}\sum_{i,j = 0}^3\eta_{i,j}\,\hat{\sigma}_{iA}\otimes\hat{\sigma}_{jB}
\label{gammae}
\end{equation}	
to generate a set of randomly perturbed operators $\hat\gamma_\epsilon$ to which the correction procedure described in Sec.\ref{Sec::generalperturbation} is applied so that the unit-trace, the constant energy, and the constant entropy constraints can be implemented and the corresponding sets of compatible randomly generated state operators $\hat{\gamma}_r^2$ be obtained and compared with the experimental ones. Then, in the spirit and intention of our construction, if the additional constraints are meaningfully related to the experimental setup, the  randomly generated state operators $\hat{\gamma}_r^2$ should be distributed in a fashion similar to the experimental state preparation of qubits on $ibmq\_ manila$.
	
\begin{figure}[!htp]
	\center
	\includegraphics[width=9cm]{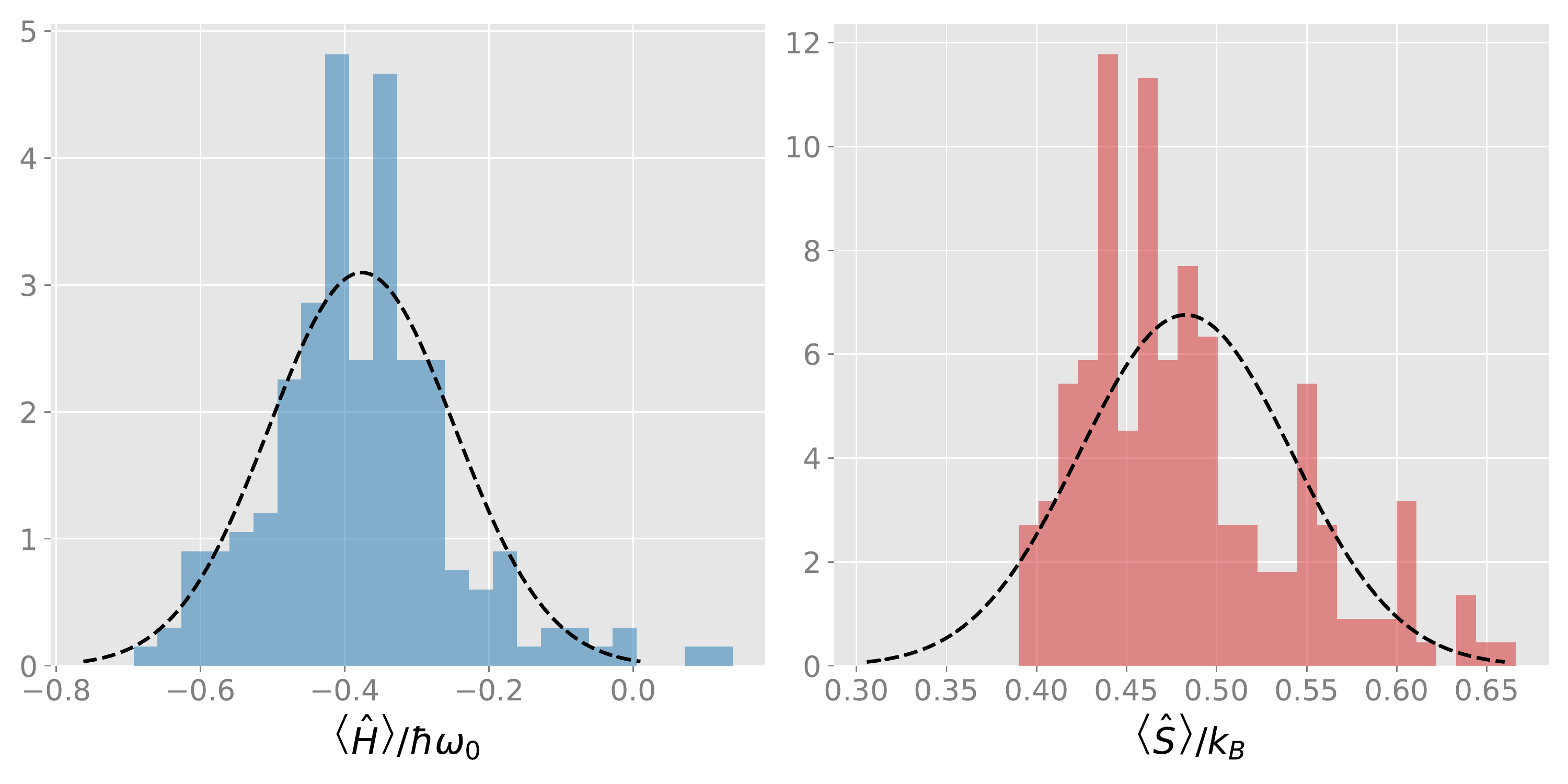}
	\caption{Histogram of the energy (left) and entropy(right) of the 200 experiments on $ibmq\_manila$.}\label{FigES}
\end{figure}

Finally, as seen in Fig. \ref{FigES}, the IBM experimental setup used here to validate the proposed perturbation procedure produces values of the energy and entropy that are normally distributed. Therefore, as a variant to the proposed procedure, it is convenient to also consider the possibility that the constraining values of energy and entropy are affected by experimental fluctuations and are, thus, not simply the fixed values of the baseline state. To  ensure that the procedure accounts for the distribution of the energy and entropy seen in Fig.  \ref{FigES}, the system of constraint equations, Eq. (\ref{Eq::Chap_4_gam_r_gam_eps_cons}), which must be solved for each $\hat{\gamma}_\epsilon$ to obtain the corresponding $\hat{\gamma}_r$, can be modified by substituting the right-hand side of the equation with randomly selected values sampled from  normal distributions with mean values and standard deviations ($\tilde E_\mu,\; \tilde E_\sigma,\; \tilde S_\mu$, and $\tilde S_\sigma$) inferred from the experimental data (e.g, see Fig. \ref{FigES}). Thus, both results for the fixed energy $\mathrm{Tr}(\rho_0 \hat H)$ and entropy  $- \mathrm{Tr}(\rho_0 \ln(\hat\rho_0))$ values of the baseline as well as normally-distributed values sampled from the normal distributions $\mathcal{N}(\tilde E_\mu, \tilde E_\sigma)$ and  $\mathcal{N}(\tilde S_\mu, \tilde S_\sigma)$, respectively, are presented in Sec. \ref{Sec::Chap_4_Results}.

\section{Results}
\label{Sec::Chap_4_Results}

The results presented here are organized into two sections. Sec.\ref{Perturbation} provides results predicted by the perturbation procedure independently of any experimental considerations, and Sec. \ref{application} couples the experimental predictions of the $ibmq\_manila$ quantum device with those of the perturbation procedure.  

\subsection{Perturbations under different constraints}
\label{Perturbation}
The results for four different constrained perturbation cases are presented in this section. Case 1 entails $\mathrm{C1}=\mathrm{Tr}(\hat{\rho}) = 1$; Case 2 C1 and $\mathrm{C2} = \mathrm{Tr}(\hat\rho \hat{H})$; Case 3 C1 and  $\mathrm{C3} = \mathrm{Tr}(\hat\rho \ \mathrm{log}\hat\rho)$; and Case 4 C1, C2, and C3. For all cases, the number of perturbed states is $N = 1000$ (as opposed to the 200 generated experimentally and presented in Sec. \ref{application}), and all random numbers $\eta_{i,j}$'s are generated from the normal distribution $\mathcal{N}(\mu=0, \sigma=0.05)$. While it is understood that this sample size is too small to thoroughly sample the entire 15-dimensional space required to fully characterize a normalized density operator of a bipartite system of two-level subsystems, it is assumed for simplicity that the trends exhibited by the perturbed states here are representative of the neighborhood of the entire 15-dimensional space. The quantities plotted are the system energy, entropy, mutual information, concurrence, and the CHSH operator maximum expectation value. In addition, the fidelity $F$ and distance measure $\theta$ of the perturbed state relative to the baseline state $\hat{\rho}_0$ are also shown. 

\begin{figure}[!htbp]
\includegraphics[width=8.5cm]{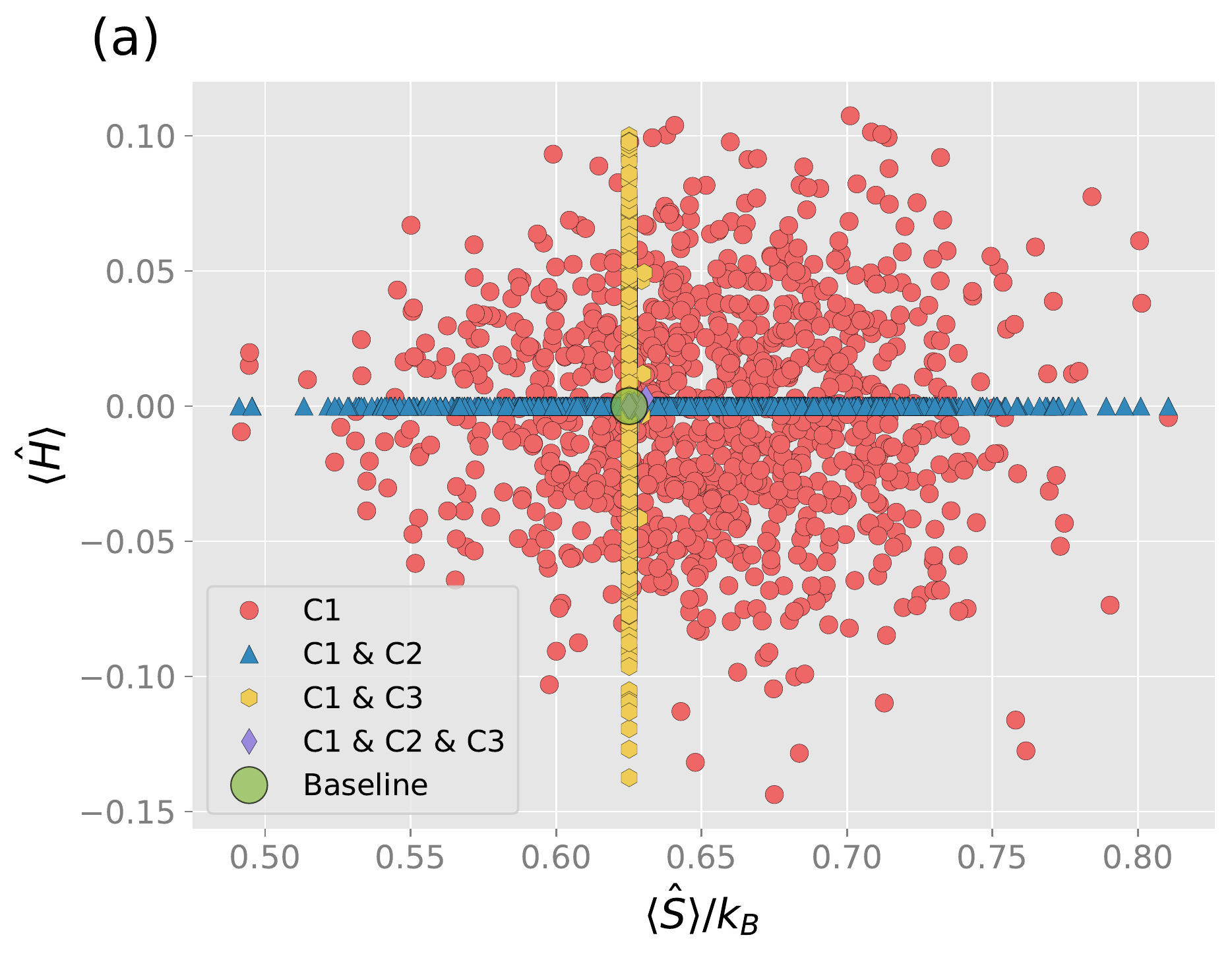}
\includegraphics[width=8.5cm]{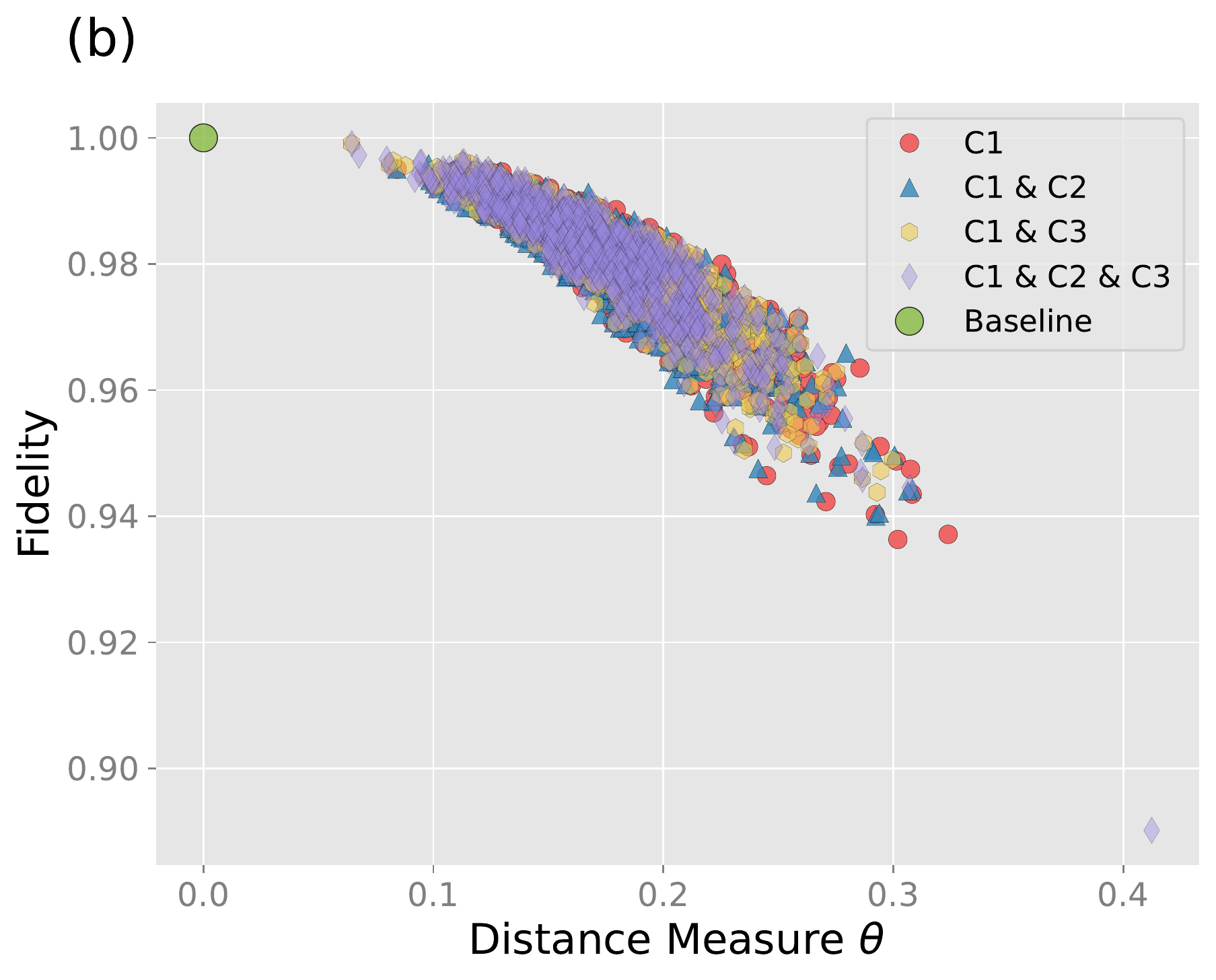}
\caption{\label{Fig3} a) Perturbed state energy-entropy diagram and b) the fidelity versus the distance measure $\theta$}.
\end{figure}

Fig. \ref{Fig3} a) shows the energy versus entropy diagram and Fig.\ref{Fig3} b) the perturbed state fidelity versus the distance measure $\theta$. In Fig. \ref{Fig3} a), it is seen that the energy and entropy of the perturbed density operators for the C1 constraint are distributed  about the baseline value of the unperturbed density operator with the heaviest concentration of values to the right of the baseline value. This concentration of values to the right is due to the fact that the set of perturbations follow a $\chi$ distribution (see Sec. \ref{appendixA}), which deviates slightly to the right of a normal distribution. In addition, as seen in this figure, the perturbed density operator values for Cases 2, 3, and 4 clearly achieve their objective of preserving the energy, the entropy, and the energy and entropy values, respectively, for the perturbed density operators. Fig. \ref{Fig3} b) shows that the distance measure $\theta$ is distributed about the mean value of 0.18 and varies $\mathrm{O}((2/3)\sigma)$ for all cases. In a similar fashion, the fidelity is distributed about the mean value 0.98 and varies an $\mathrm{O}((1/5)\sigma)$ for all cases.

\begin{figure}[!htbp]
\center
\hspace*{-0.6cm}\includegraphics[width=9.5cm]{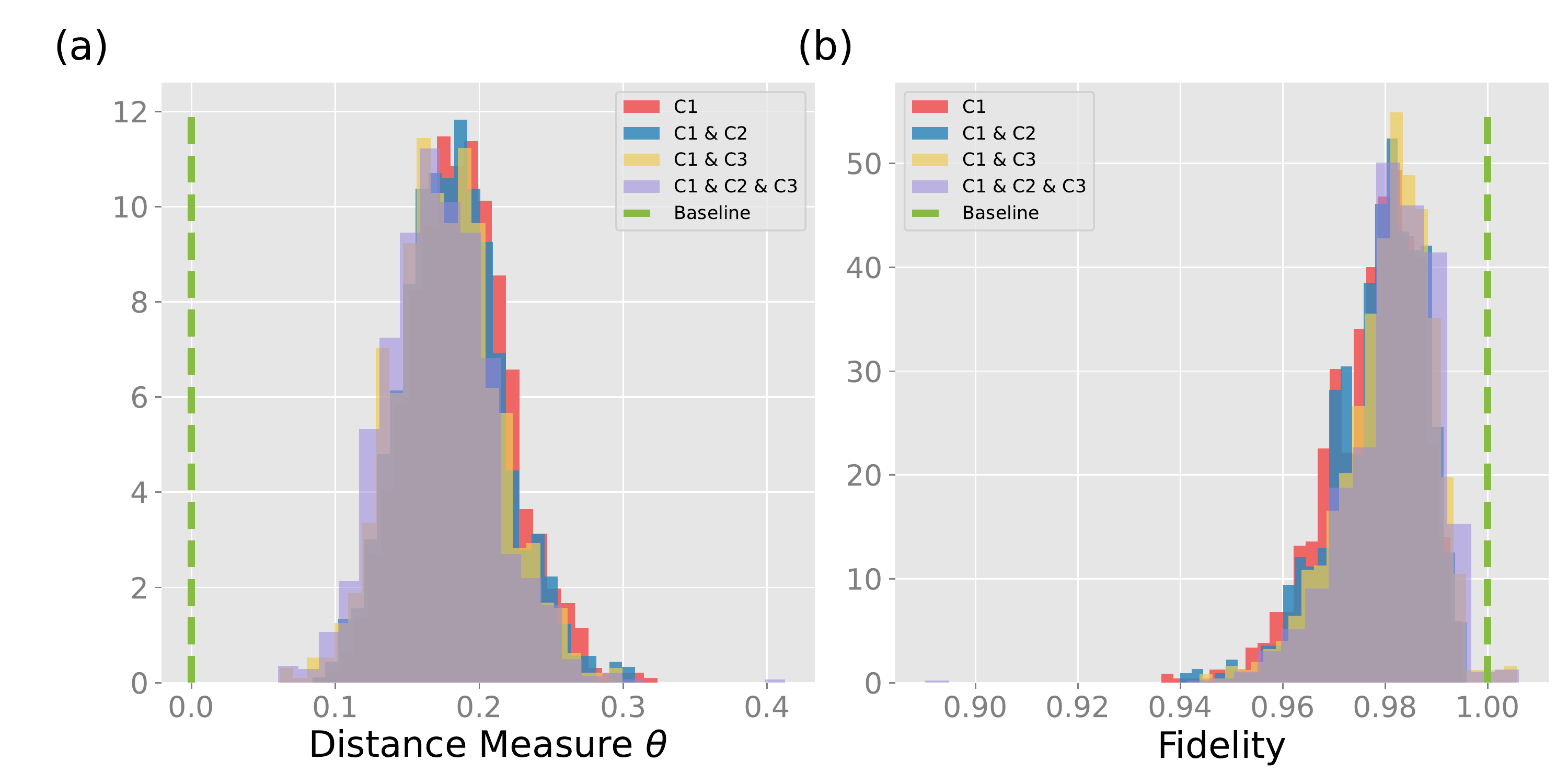}
\caption{\label{FigHist1} a) Histogram of the perturbed states relative to the distance measure $\theta$ and b) histogram of the perturbed states relative to the fidelity. The y-axis for both cases show the probability density.}
\end{figure}

Histograms of the distance measure $\theta$ and the fidelity of Fig. \ref{Fig3} are shown in Figs. \ref{FigHist1} a) and b), respectively. Here, the distance measure is distributed according to a $\chi$ distribution, while the fidelity follows a $\chi$-square distribution. The y-axis for both cases in Fig. \ref{FigHist1} show the probability density, this means that each bin in the histogram displays the bin's raw count divided by the total number of counts times the bin width sothat the area under each histogram integrates to one. The same is true for all of the histograms in the succeeding figures.

\begin{figure}[!htbp]
\includegraphics[width=9cm]{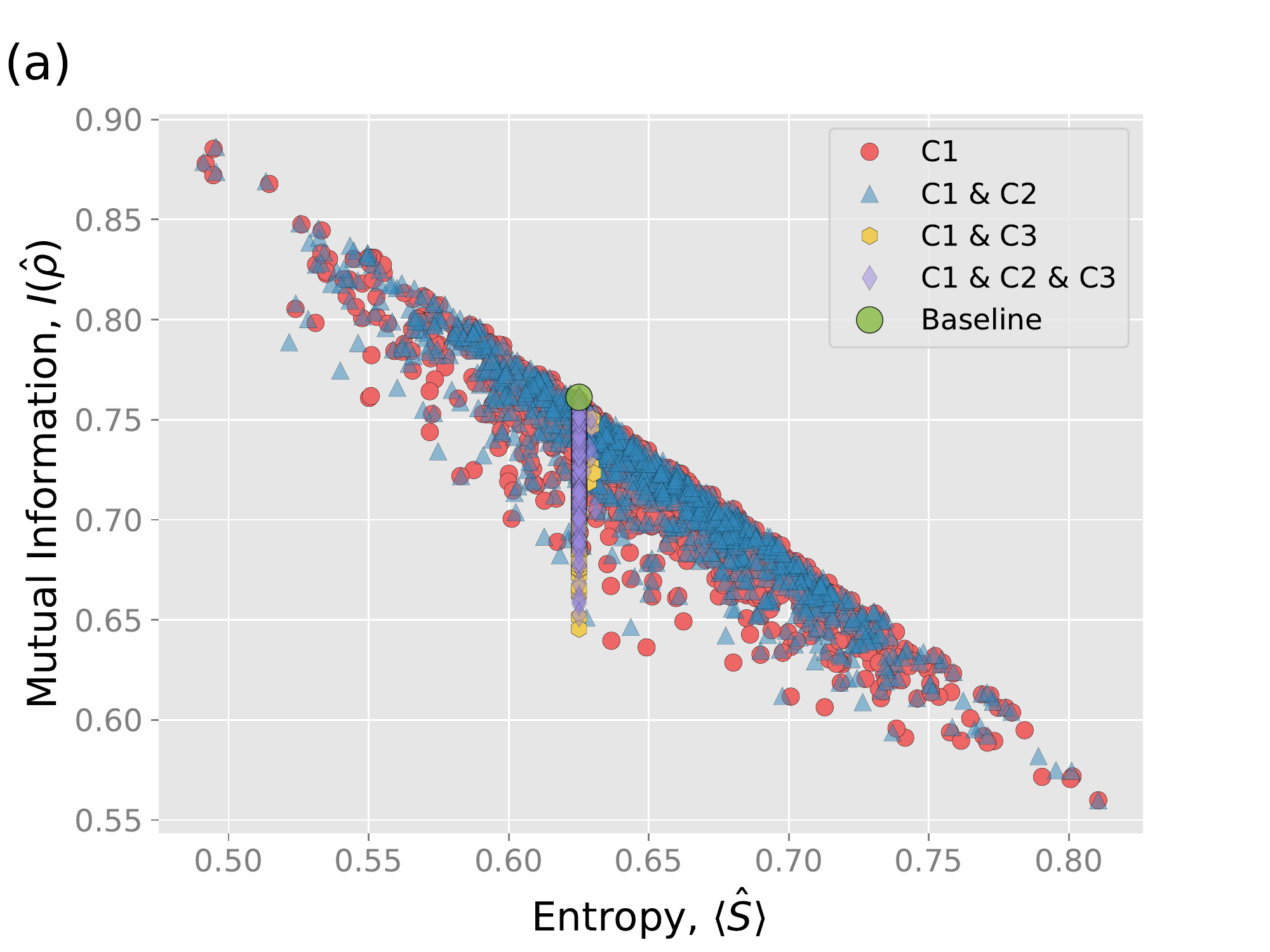}
\includegraphics[width=9cm]{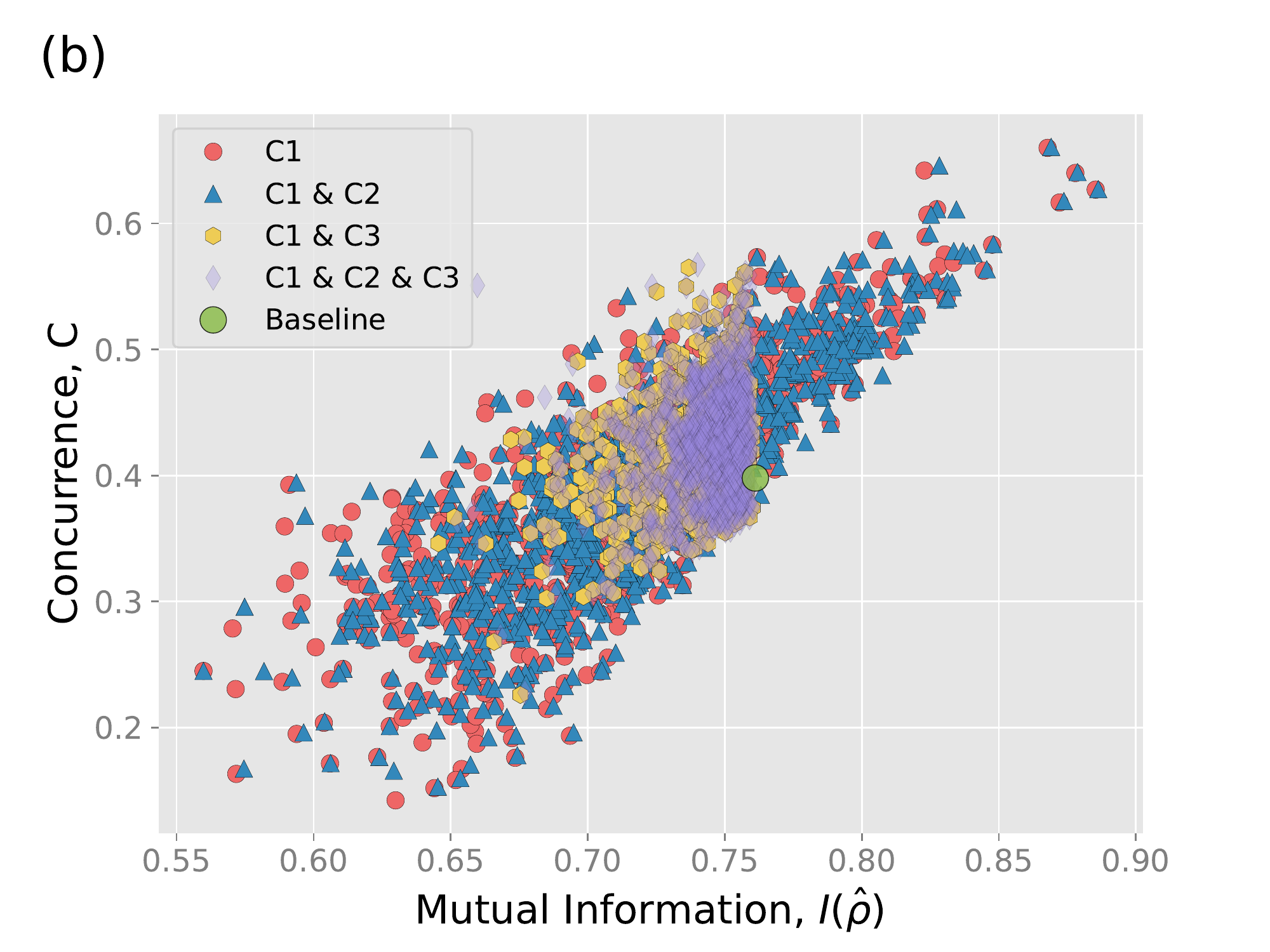}
\caption{\label{Fig5} Perturbed state mutual information versus the entropy and the concurrence versus the mutual information.}
\end{figure}

Fig. \ref{Fig5} a) shows the mutual information of the perturbed state versus the entropy, and Fig. \ref{Fig5} b) provides the perturbed state concurrence versus the mutual information. Fig. \ref{Fig5} a) indicates that the perturbed state mutual information decreases approximately linearly as the entropy increases for all constrained cases except for those where the entropy is constant. For theses cases, the whole set of mutual information values lie vertically directly below the baseline value. Furthermore, the points are distributed unevenly from the baseline, something that can be explained by the $\chi$ distribution that is obtained from a set of normally distributed random variables as is explained in Sec.\ref{appendixA}. In Fig. \ref{Fig5} b), it is seen that the mutual information varies an $\mathrm{O}(\sigma)$ from the baseline value for Cases 1 and 2 and an $\mathrm{O}((1/3)\sigma)$ for Cases 3 and 4, while the concurrence varies an $\mathrm{O}((5/3)\sigma)$ from the mean value for Cases 1 and 2 and an $\mathrm{O}(\sigma)$ for Cases 3 and 4. Also, as can be seen in the figure, there is a clear trend between the value of the mutual information and that of the concurrence of a given state since as expected for a state that decreases the mutual information with the baseline there is also a loss of entanglement. 

\begin{figure}[!htbp]
\center
\hspace*{-0.6cm}\includegraphics[width=9.5cm]{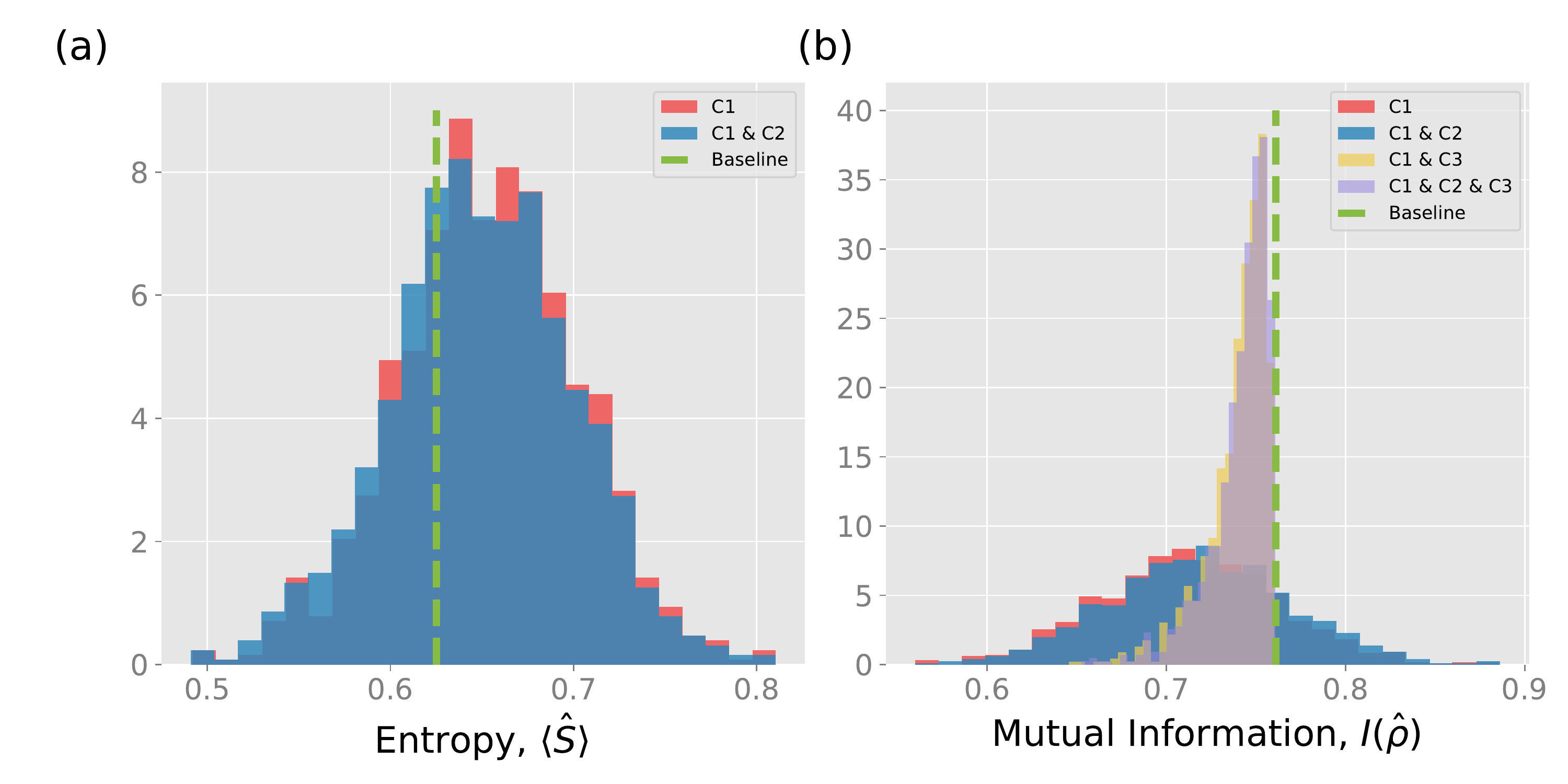}
\caption{\label{FigHist2} a) Histogram of the perturbed states relative to the entropy $\langle \hat S \rangle$ and b) histogram of  the perturbed states relative to the mutual information. The y-axis for both cases show the probability density.}
\end{figure}

Figs. \ref{FigHist2} a) and b) show the histograms for the entropy and the mutual information of Fig. \ref{Fig5}. As is seen in Fig. \ref{FigHist2} a), the entropy is distributed (as it was in Fig. \ref{Fig3} a)) slightly to the right of the baseline indicating that values with higher entropy are more prone to occur owing to the perturbation characteristics. For the the mutual information, Fig. \ref{FigHist2} b) shows that Cases 1 and 2 are normally distributed with a mean value to the left of the baseline, indicating that the perturbation reduces the entanglement between the two subsystems. For cases 3 and 4, the mutual information follows a $\chi$-square distribution.

\begin{figure}[!htbp]
\includegraphics[width=9cm]{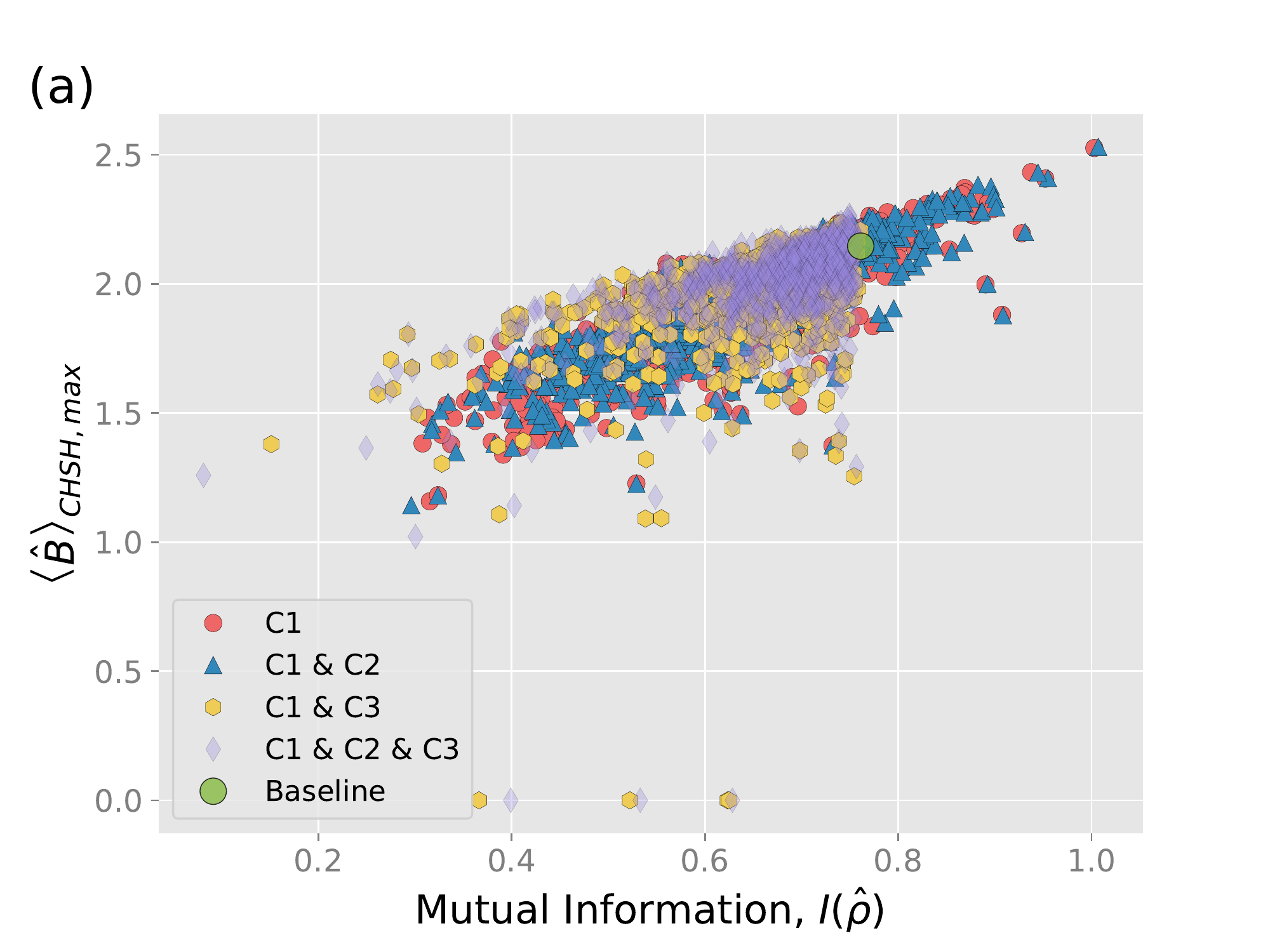}
\includegraphics[width=9cm]{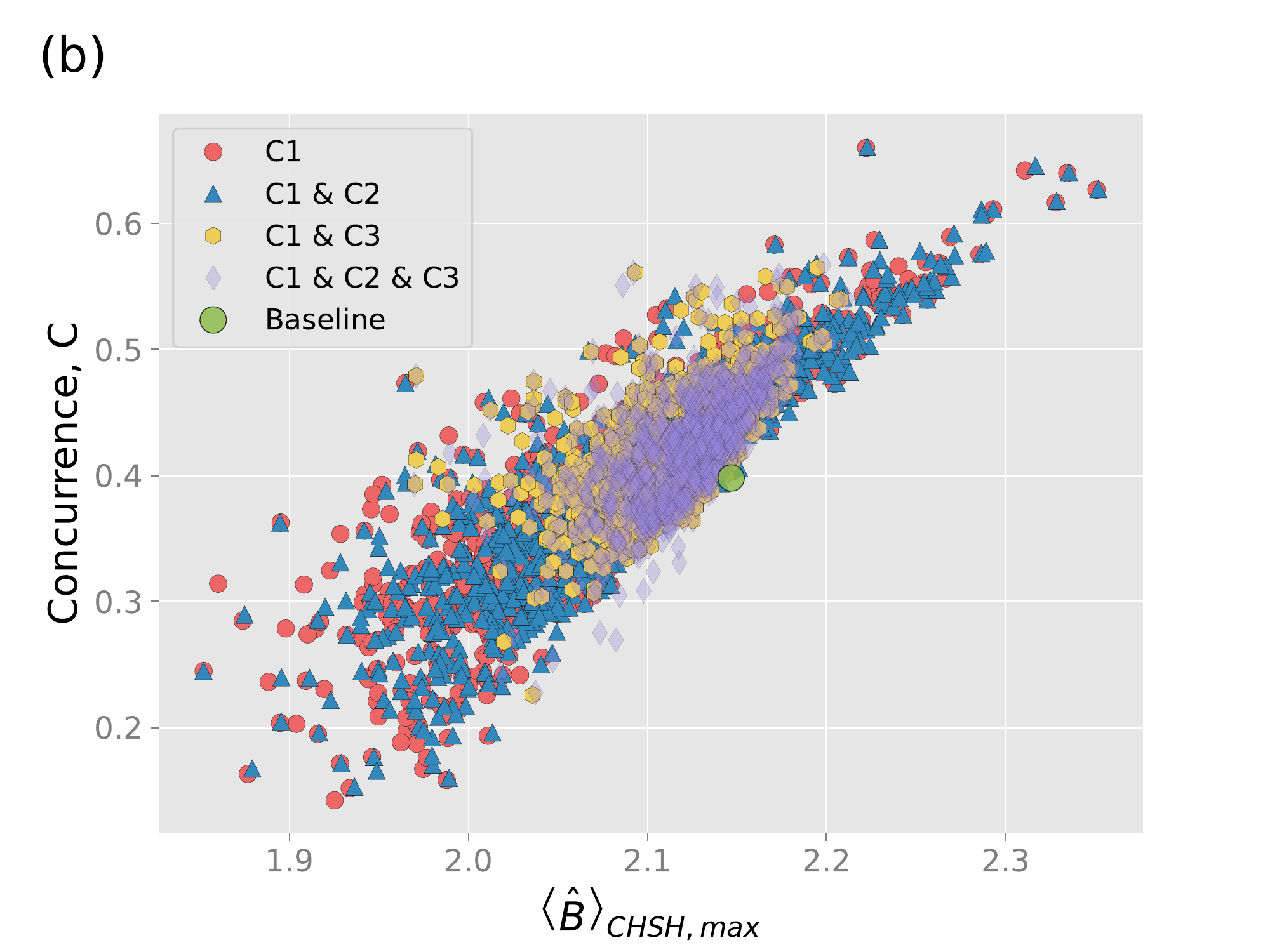}
\caption{\label{Fig7}a) Perturbed state maximum CHSH operator expectation value versus the mutual information and b) the concurrence versus the maximum CHSH operator expectation value.}
\end{figure}

Fig. \ref{Fig7} a) shows the maximum CHSH operator expectation value versus the mutual information of the perturbed state, and Fig. \ref{Fig7} b) provides the perturbed state concurrence versus the maximum CHSH operator expectation value. Fig. \ref{Fig7} a) indicates that the increase in the perturbed state maximum CHSH operator expectation value is proportional to the mutual information increase for Case 1 and 2, which is again expected since both aim to quantify the entanglement of a system's state. However, for the constrained cases that include C3 (i.e., Cases 3 and 4), there is a restriction in the mutual information that can be reached with a value below that of the baseline.

As to Fig. \ref{Fig7} b), it shows that the concurrence and the maximum CHSH operator expectation value vary asymmetrically from the baseline. Thus, those values with less concurrence and maximum CHSH expectation values are prone to occur and there is a clear trend between the value of the mutual information and the value of the concurrence of a given state. In both Figs. \ref{Fig7} a) and b), the extent of the spread of the perturbed state values varies and is in particular greatly impacted by the entropy restriction. The spread on the concurrence and mutual information are as indicated above in the discussion surrounding Fig. \ref{Fig5}, while that for the maximum CHSH expectation values vary an $\mathrm{O}((3/2)\sigma)$ from the mean value for Cases 1 and 2 and almost an $\mathrm{O}((2/3)\sigma)$ for Cases 3 and 4.  

\begin{figure}[!htbp]
\center
\hspace*{-0.6cm}\includegraphics[width=9.5cm]{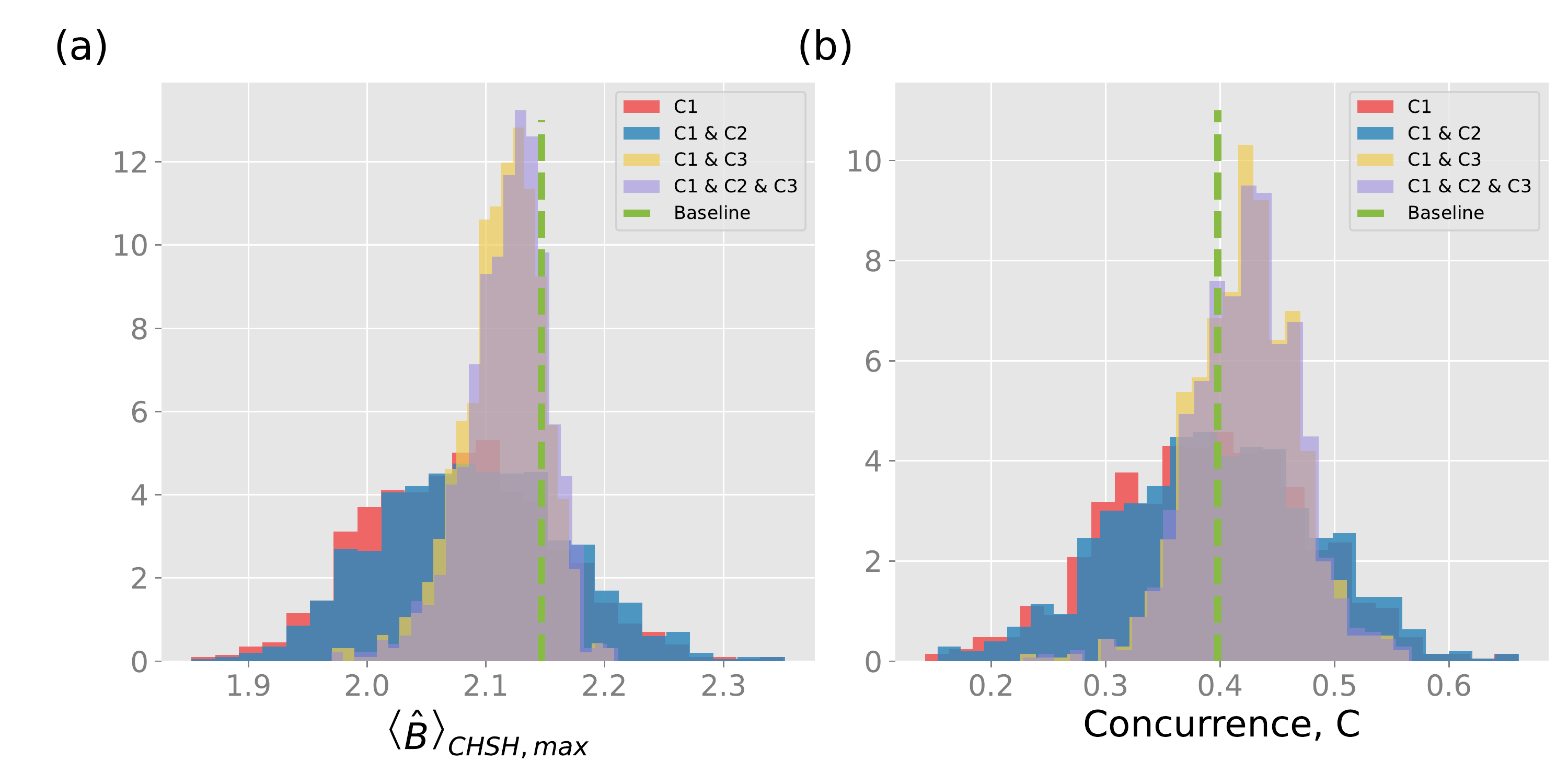}
\caption{\label{FigHist3} a) Histogram of the perturbed states relative to the maximum CHSH operator and b) histogram of the perturbed states relative to the concurrence. The y-axis for both cases show the probability density.}
\end{figure}

The histograms for the perturbed state distribution relative to the maximum CHSH operator and the concurrence of Fig. \ref{Fig7} b) are shown in Figs. \ref{FigHist3} a) and b), respectively. For the case of the CHSH operator, values to the left of the baseline are more likely to occur (see Fig. \ref{FigHist3} a)),and the C3 constraint considerably reduces  the spread of the points. Something similar occurs with the concurrence, except that in this case, values to the right of the baseline are more likely (see Fig. \ref{FigHist3} b)). It is also interesting to note that the C3 constraint affects the entanglement so that there is a greater probability of higher values of entanglement when the C3 constraint is present than when it is not.

\subsection{Applications of the perturbed state generation to the IBM transmon device}
\label{application}
In this section, the simulation results for 200 Bell states prepared on $ibmq\_manila$ as explained in Section \ref{setup} are presented. Experimental fluctuations in the constraining values of the energy and entropy are taken into account in these simulations using random values of the energy and entropy sampled from the experimentally generated normal distributions $\:\mathcal{N}(\tilde E_\mu, \tilde E_\sigma)$ and  $\mathcal{N}(\tilde S_\mu, \tilde S_\sigma)$ given in Fig. \ref{FigES}. To generate the set of randomly perturbed states, 200 randomly generated values for each $\eta_{i,j}$ are also determined from normal distributions with $\mathcal{N}(\mu_{ij} = 0, \sigma_{ij}=0.05)$. However, simply using these to generate the random set of 200 perturbed density operators with a distribution of energies $\tilde E_\mu$'s and entropies $\tilde S_\mu$'s similar to that for the density operators resulting from the experiments is problematic in that the $\eta_{i,j}$ values are effectively uncorrelated, while those for the experiments are not.

\begin{figure}[]
\includegraphics[width=9cm]{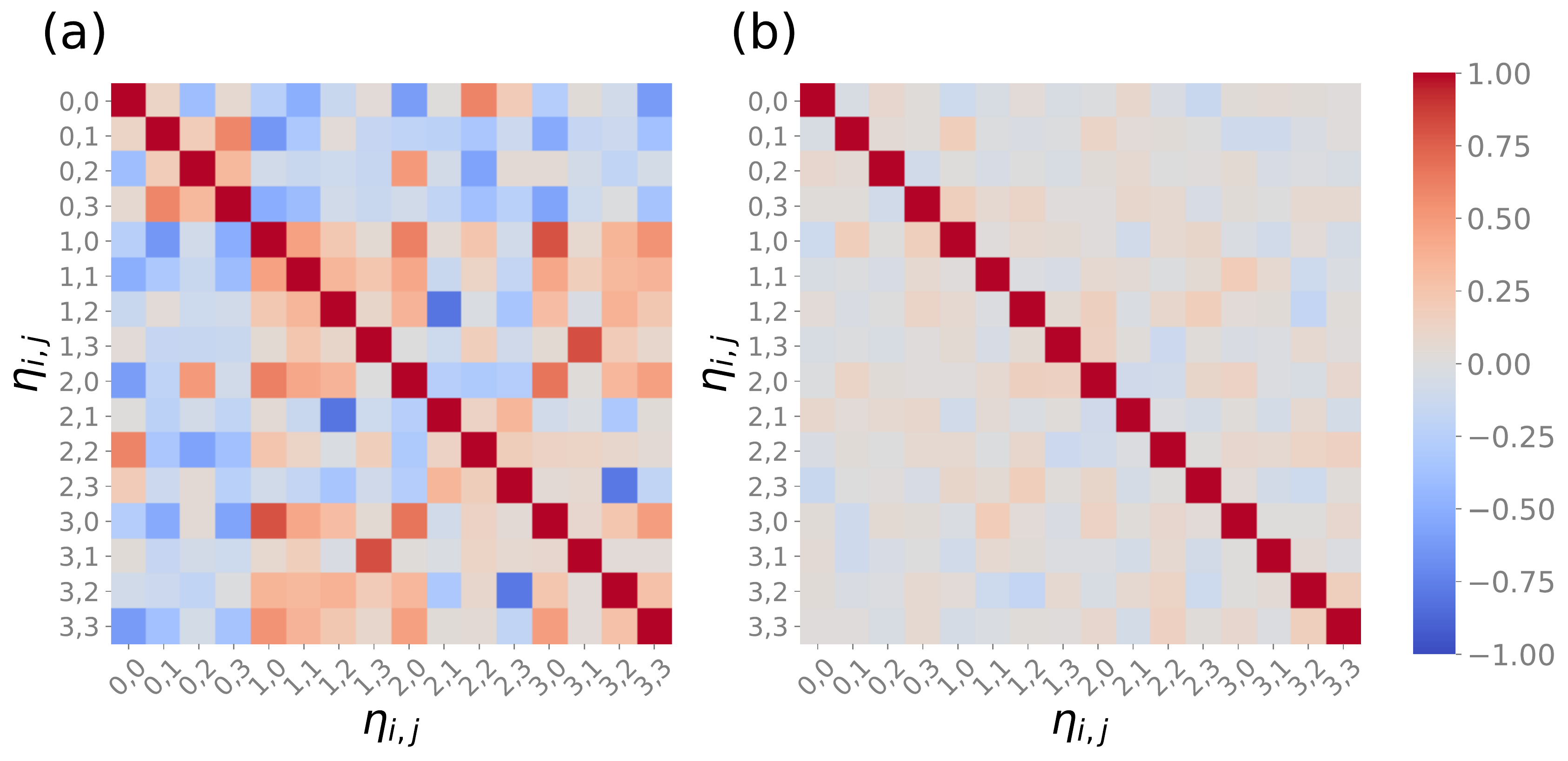}
\caption{\label{FigCorr} Correlation diagram of a) the experimental $\eta_{i,j}$values and b) the simulated $\eta_{i,j}$ values.  Here, +1 and -1 indicates an exact linear relationship between values while a value of 0 indicates non-correlated values.}
\end{figure}

This is seen clearly in Fig. \ref{FigCorr} a) where each square represents a Pearson correlation value between two experimental $\eta_{i,j}$ with a value of 1 or -1 indicating a linear correlation and a value of 0 no correlation. In contrast and as expected, Fig. \ref{FigCorr} b) shows a Pearson correlation value close to 0 for all of the randomly generated $\eta_{i,j}$ values.

\begin{figure}[]
	\includegraphics[width=9cm]{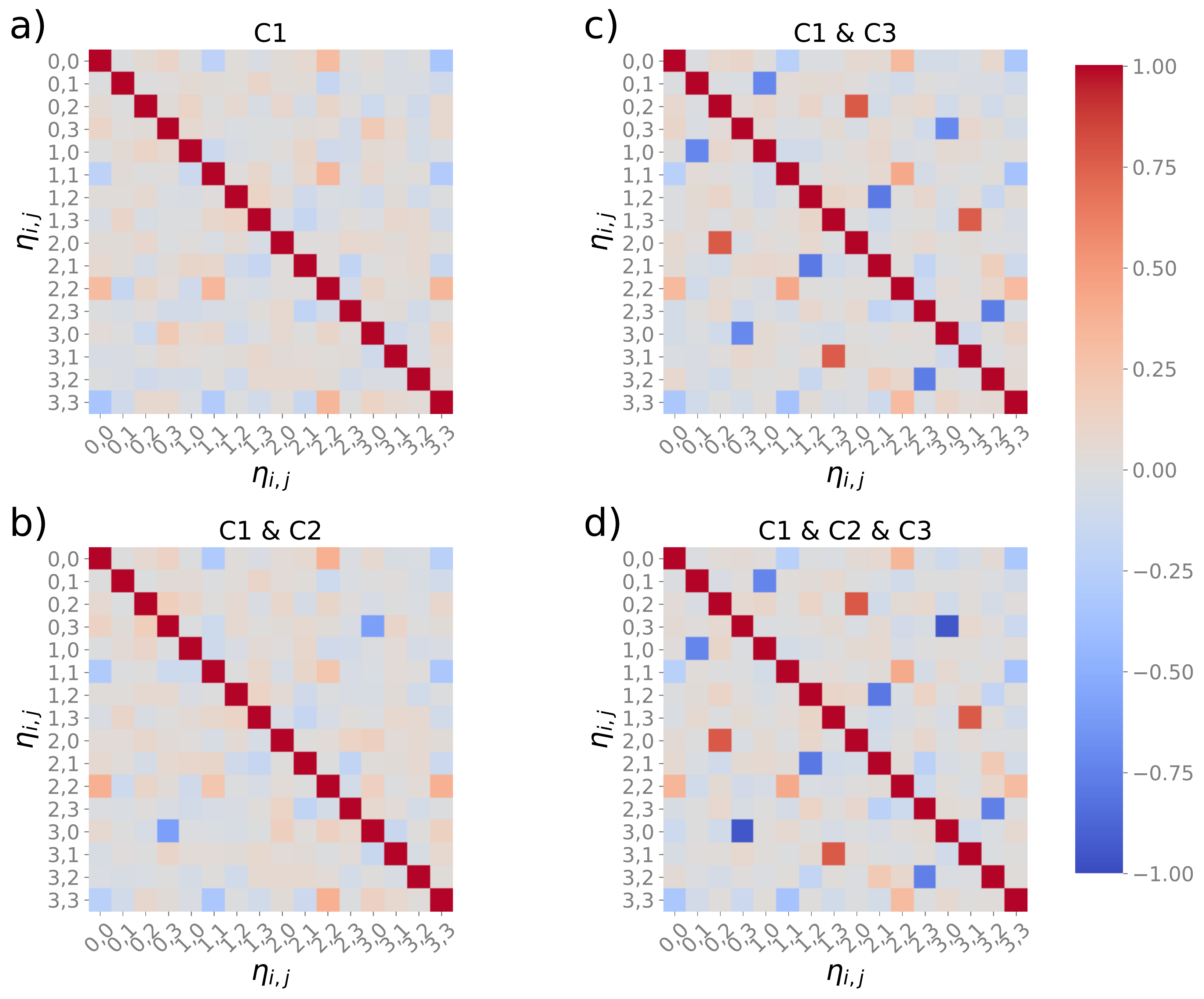}
	\caption{\label{Figcorr1} Correlations between the a) $\eta^{C1}_{i,j}$, b) $\eta^{C1, C2}_{i,j}$, c) $\eta^{C1, C3}_{i,j}$, and d) $\eta^{C1, C2, C3}_{i,j}$ values after applying the different constraints.}
\end{figure}

However, as seen in Fig. \ref{Figcorr1} a), once the first constraint, C1, is applied to the random generation of $\hat{\gamma}_\epsilon$ to transform it from a non-Hermitian operator to the Hermitian operator $\hat{\gamma}_R$, correlations between the $\eta_{i,j}$ appear.  These increase with the addition of the constraint C2, but, mainly with the addition of constraint C3, as seen in Figs. \ref{Figcorr1} b) to d) where it is noted that the density operator generated without the C3 constraint is less correlated.

\begin{figure}[]
\includegraphics[width=9cm]{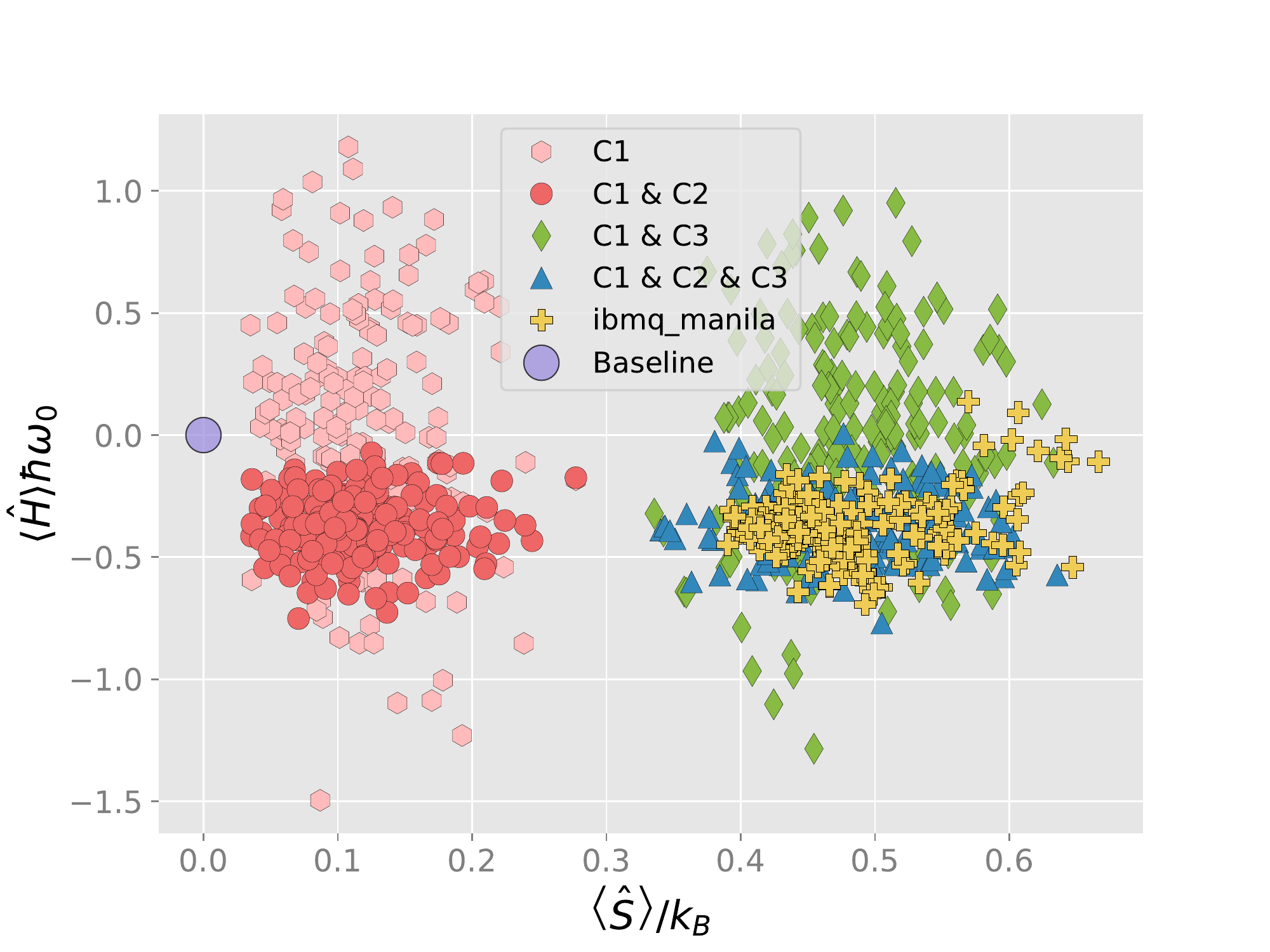}
\caption{\label{Fig8}Energy-entropy diagram of the simulation of the ibmq\_manila device based on the $\eta_{i,j}$ values using the normal distributions $\mathcal{N}(\tilde E_\mu, \tilde E_\sigma)$ and $\mathcal{N}(\tilde S_\mu, \tilde S_\sigma)$ for the energy and entropy, respectively. The yellow crosses are the 200 experimental values.}
\end{figure}

Fig. \ref{Fig8} shows the energy-entropy distribution for the perturbed density operators generated with different sets of constraints. The yellow crosses in this figure represent the density operators resulting from the 200 experiments. These experimental states are distributed between an entropy of about 0.4 and 0.7 and, for the most part, at an energy a little below the baseline energy when the energy is constrained. As can be seen, the states generated for Cases 1, 2, and 3 result in much broader spreads, respectively, in terms of the energy and entropy. This contrasts with the spread of states generated with Case 4, which closely approximates that for the experimental results from $ibmq\_manila$. As a result, the following figures, Figs. \ref{Fig9} to \ref{Fig11}, focus on Cases 2 to 4 and exclude Case 1 since its spread is the least representative of that for the experiments. 

Fig. \ref{Fig9} shows a) the energy-entropy distribution and b) the fidelity versus the distance measure. The figures at the top and on the right side of each of these figures represent histograms of the number of simulated and experimental states. As can be seen in Fig. \ref{Fig9} a), all the energy and entropy values for Case 4 are comparable to those for the experiment (the yellow crosses). For Case 3 this is only true for the entropy values, while for Case 2 neither the entropy nor the energy values match the experimental. In contrast, Fig. \ref{Fig9} b), shows that variations in both distance measures for Cases 3 and 4 match those of the experiment well. This is not true for Case 2.

\begin{figure}[]
\includegraphics[width=8.5cm]{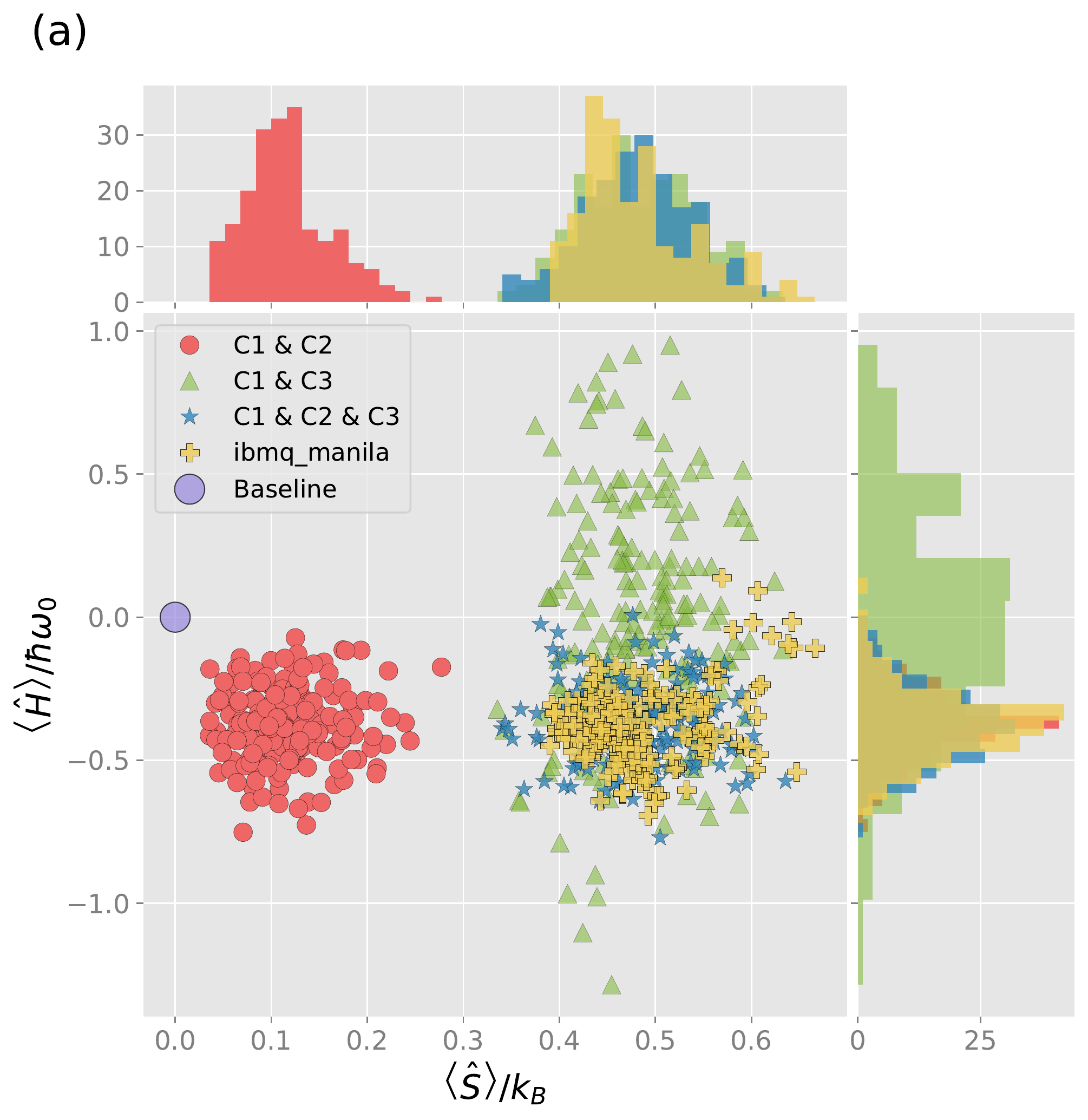}
\includegraphics[width=8.5cm]{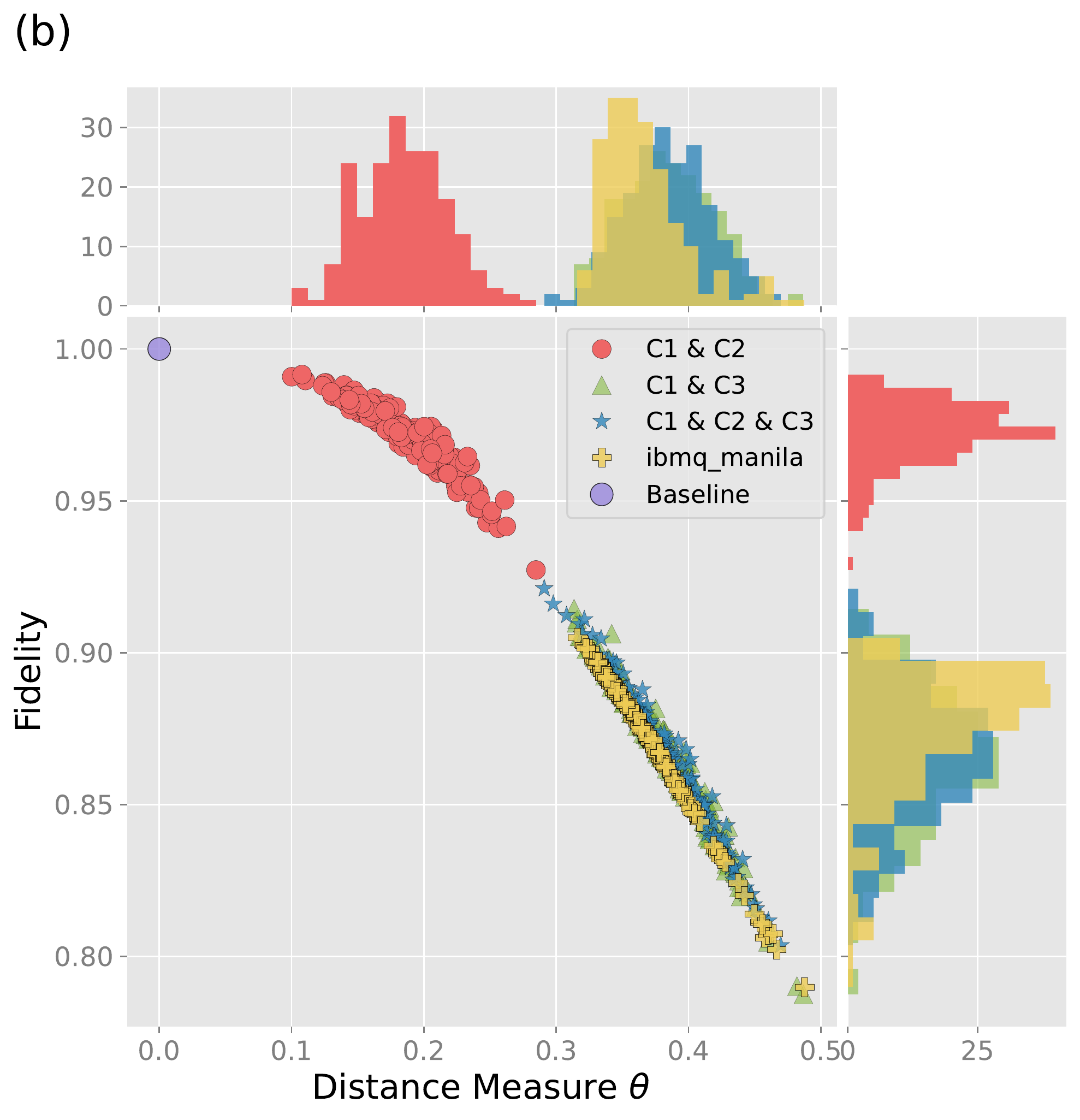}
\caption{\label{Fig9} a) Energy-entropy diagram of the simulation of the ibmq\_manila device based on the $\eta_{i,j}$ values using the normal distributions $\mathcal{N}(\tilde E_\mu, \tilde E_\sigma)$ and $\mathcal{N}(\tilde S_\mu, \tilde S_\sigma)$ for the energy and entropy, respectively. The yellow crosses are the 200 experimental values; b) the fidelity versus the distance measure $\theta_d$ for these states.}
\end{figure}

Fig. \ref{Fig10} a) shows the entropy versus mutual information distribution for states generated with Cases 2 to 4. As in Fig. \ref{Fig9}, histograms of the number of states appear at the top and on the right side. As seen, values for the mutual information for Case 2 are higher than those for the experiment and Cases 3 and 4 and span a lower range of the entropy. In contrast, the values for the entropy and mutual information for Cases 3 and 4 overlay the experimental values (yellow crosses) fairly well with the commonality being that both cases apply the entropy constraint. A similar behavior is observed with the values of the concurrence and mutual information seen in Fig. \ref{Fig10} b). Case 2 again lies outside the experimental range, while Cases 3 and 4 overlay this range. The commonality once more is the entropy constraint.

\begin{figure}[]
\includegraphics[width=8.5cm]{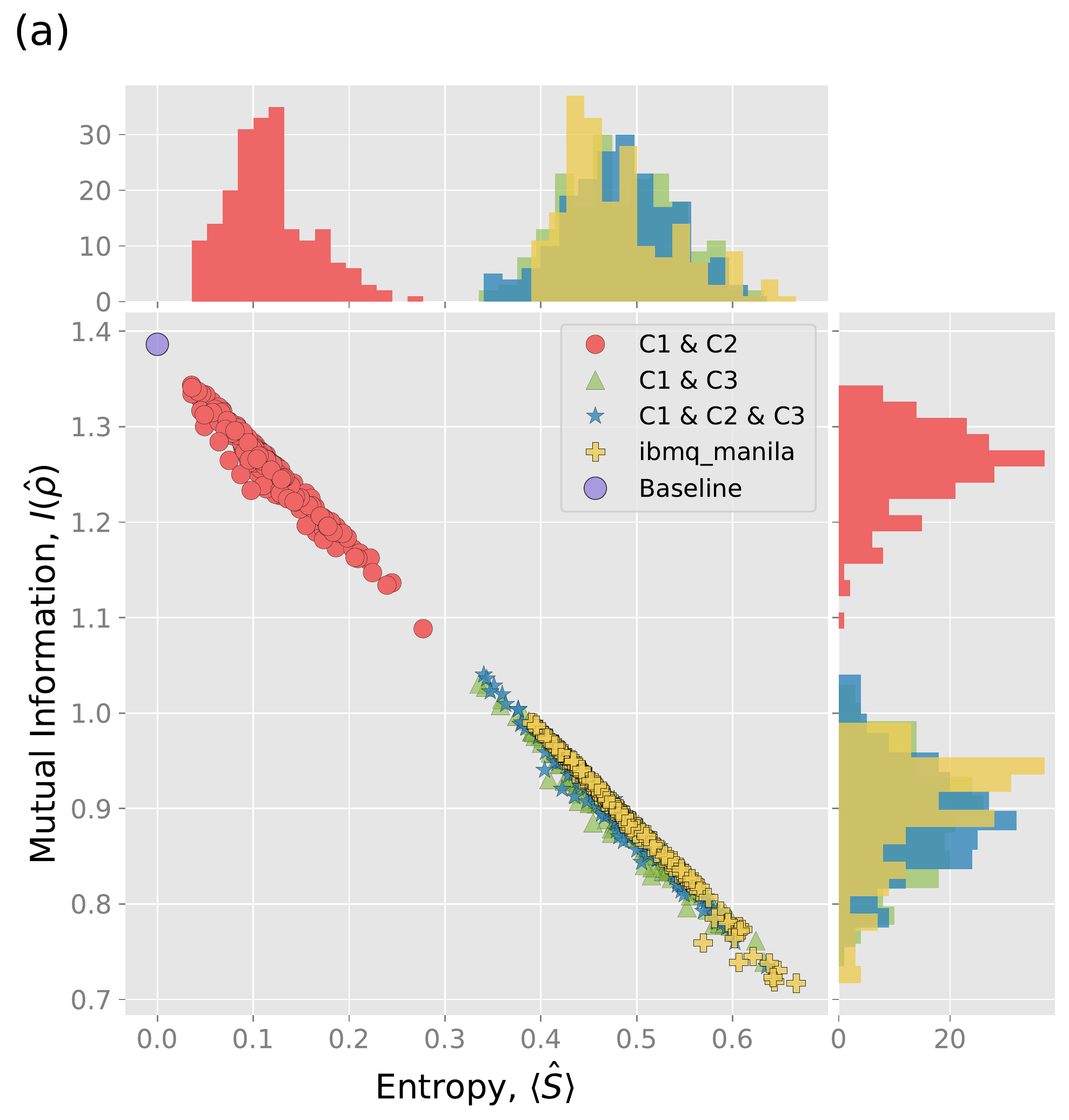}
\includegraphics[width=8.5cm]{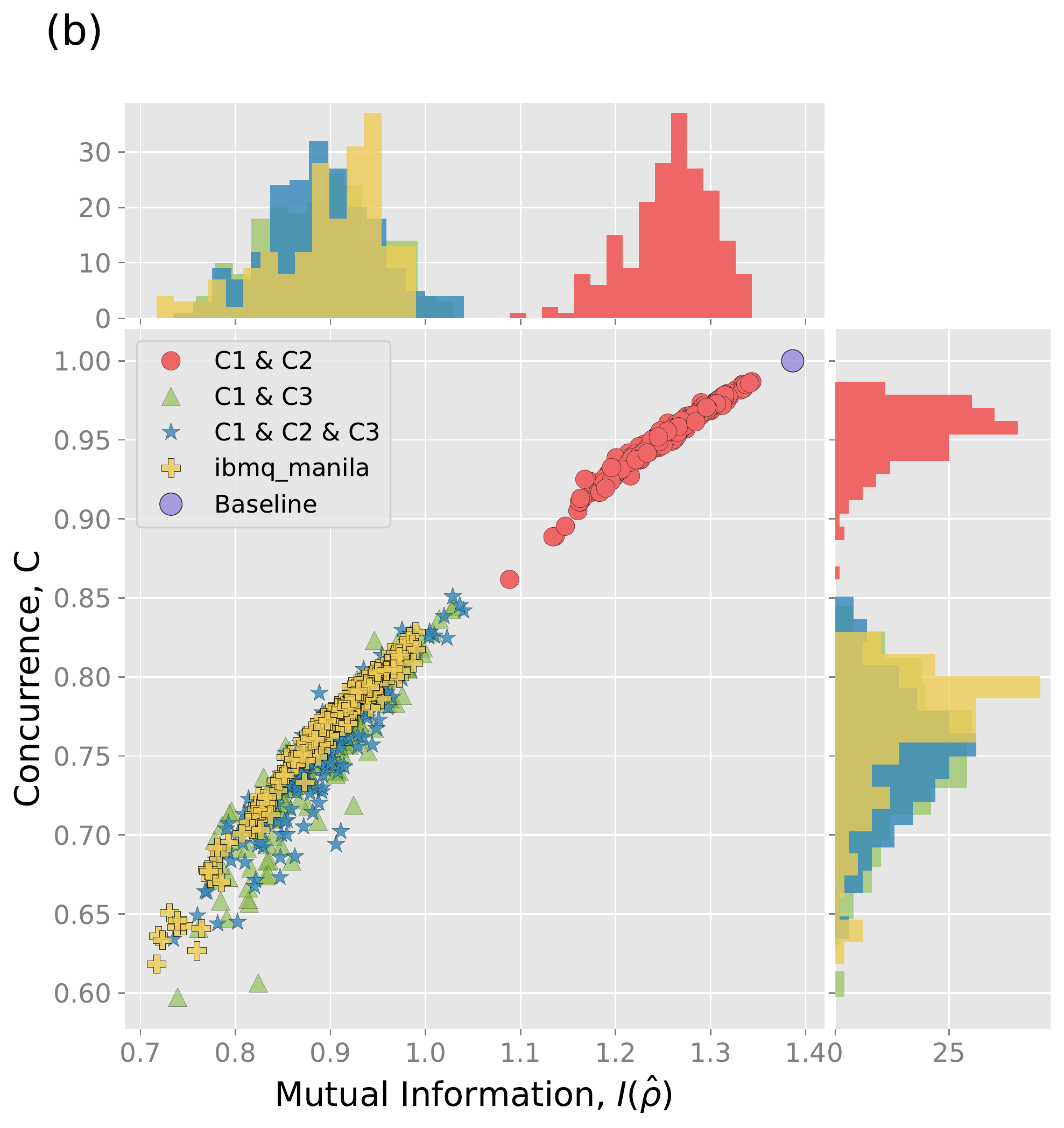}
\caption{\label{Fig10}  a) Mutual information versus the entropy and b) the concurrence versus the mutual information for the experimental and randomly generated perturbed Bell states.}
\end{figure}

Finally, Fig. \ref{Fig11} a) shows the maximum CHSH expectation values versus the mutual information distribution and Fig. \ref{Fig11} b) the concurrence versus the maximum CHSH expectation values. Histograms again are at the top and on the right. As before, the states generated for Cases 3 and 4 provide the best approximations to the experimental results, demonstrating the importance of the entropy constraint.

\begin{figure}[]
\includegraphics[width=8.5cm]{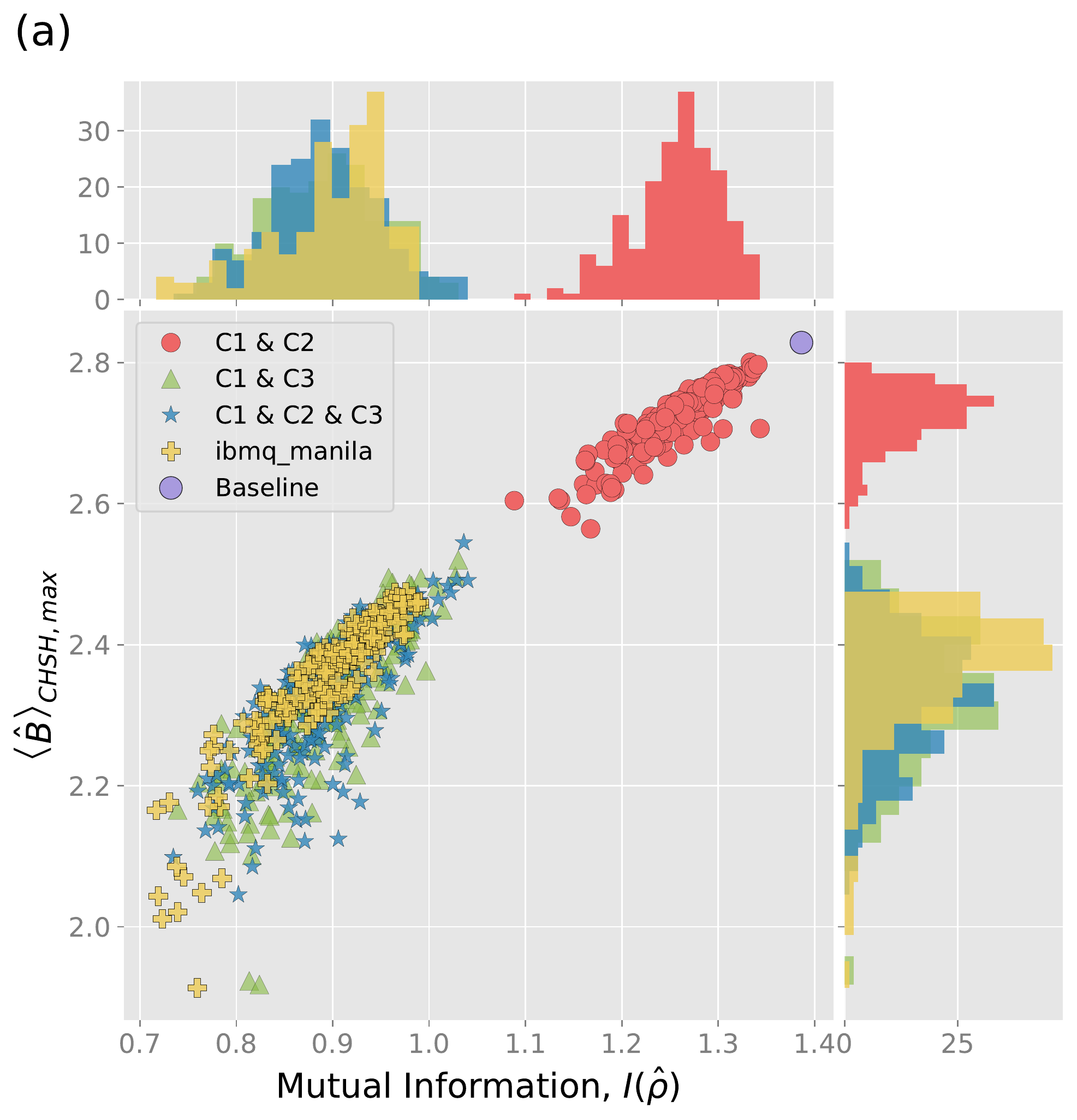}
\includegraphics[width=8.5cm]{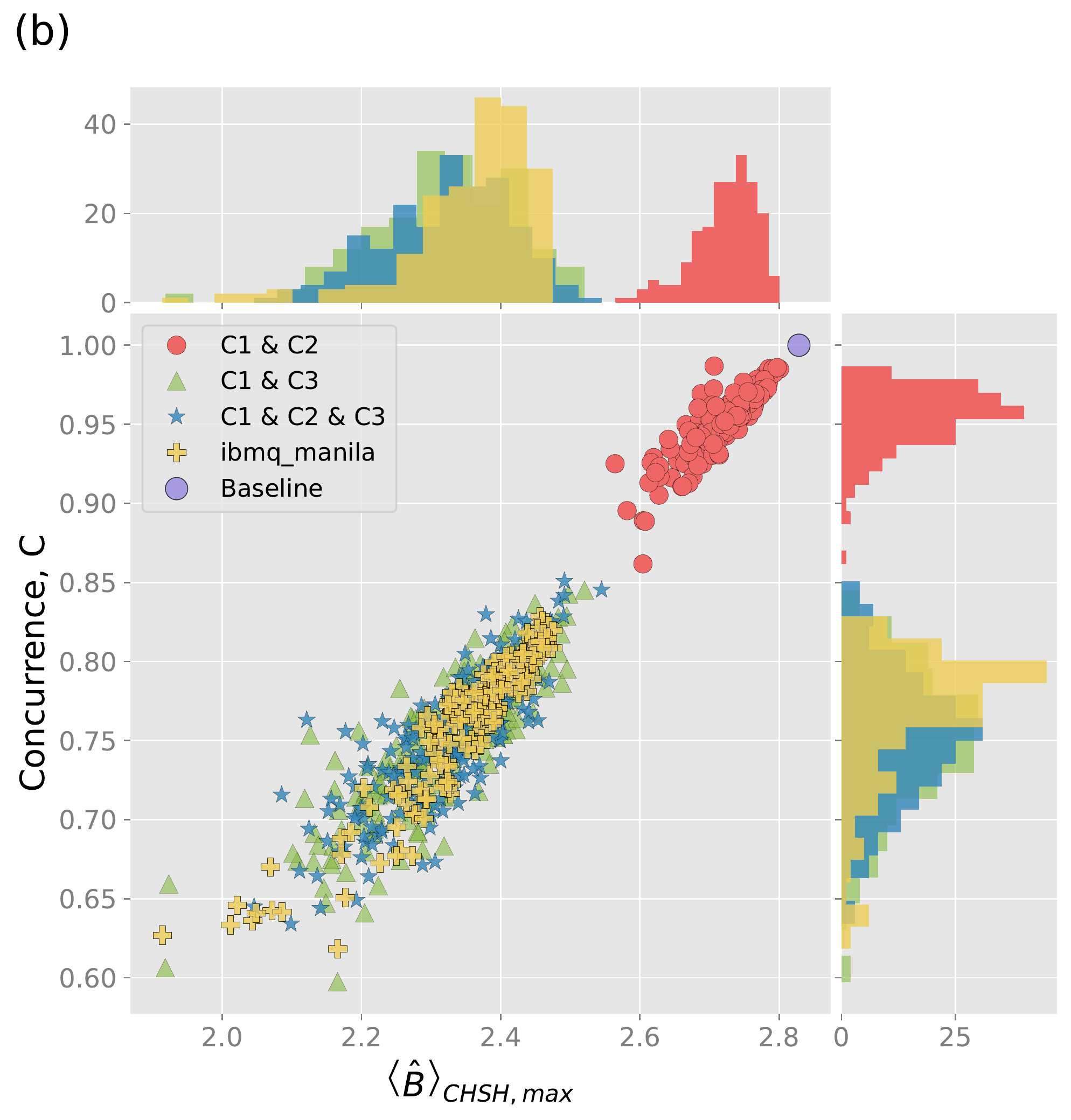}
\caption{\label{Fig11} a) Maximum CHSH operator expectation value versus the mutual information and b) the concurrence versus the maximum CHSH operator expectation value for the experimental and the randomly generated perturbed Bell states.}
\end{figure}

\section{Conclusions}
\label{Sec::Chap_4_Conclusion}

This paper presents a general perturbation method for randomly creating perturbed states that are a specified distance $\theta$ away from the original baseline state, with $\theta$ defined in Section \ref{SSec::Chap_4_Dist_Entang_Meas}. The method furthermore permits arbitrary sets of constraints to be applied to the expectation values of the perturbed states and is illustrated by using combinations of energy and entropy constraints to generate perturbations of a Bell diagonal state, a so-called maximally entangled state, as well as perturbations of experimentally prepared Bell states. The effects of the various types of perturbations on the entanglement characteristics of the resulting perturbed states are presented. 

A first observation is that, for the simulated values when experimental fluctuations are not considered, the constant entropy constraint has a major effect on the distribution of the perturbed state entanglement characteristics. This suggests that the entanglement of two systems is closely correlated to the composite system entropy and that, for example, as the system entropy increases, the entanglement decreases. In fact, the magnitude of the change of the mutual information from the baseline state is directly proportional to the magnitude of change in the entropy from the baseline state. This observation is, of course, likely dependent on the choice of baseline state (i.e., in this case, a Bell diagonal state), and further work should be done to examine this trend for a broader class of states.  

A second observation is that the differences in the distance measures of the perturbed states from the base state do not greatly depend on which constraints are applied. For all the cases, the distance measure $\theta$ varies an $\mathrm{O}((2/3)\sigma)$ away from the mean value, while the fidelity varies an $\mathrm{O}((1/5)\sigma)$. The principal difference between these measurements of the "closeness" between two states is the distribution that best fits the spread of perturbed states. For the case of the distance measure $\theta$, it is a $\chi$ distribution, while for the fidelity it is a $\chi$-square distribution.
 
Regarding  the simulated values when experimental fluctuations are taken into account, the first observations is that, without both the energy and entropy constraints, determining the simulation $\eta^n_{ij}$ coefficients based on the experimentally generated normal distributions is problematic since these coefficients are effectively uncorrelated. Including the constraints introduces correlations similar to what is seen in the experimentally generated $\tilde{\eta}^n_{ij}$ coefficients.  Without these correlations, generating simulated perturbed density operators comparable to those generated by the experiments is not possible. 

Another observation is that using the perturbation method proposed here, a large number of density operators can be constructed with a limited number of experiments to provide the necessary statistics. This is advantageous for theoretical models that need a large sampling of density operators. To generate all the needed density operators experimentally would be extremely time consuming and costly. In addition, this approach can be extended to study the behavior of NISQ devices from a theoretical perspective and to characterize the devices in terms of the experimental $\eta_{i,j}$. Further investigation in this direction, however, is beyond the scope of the present paper.

Finally, future extensions of this work include the development of a broader set of constraints to apply in the perturbation procedure (e.g., a constant concurrence constraint) as well as a more in-depth statistical analysis to better understand how certain types of perturbations change various system properties and which correlations between variables are strongest.

\begin{acknowledgments}
\vspace{-10pt}
J. A. Monta\~nez-Barrera thanks the National Council of Science and Technology (CONACyT), Mexico, for his Assistantship No. CVU-736083. 

We acknowledge use of the IBM Q for this work. The views expressed are those of the authors and do not reflect the official policy or position of IBM or the IBM Q team.

Finally, R. T. Holladay would like to thank the U.S. Department of Defense for its Science, Mathematics, and Research for Transformation (SMART) scholarship.
\end{acknowledgments}

\bibliography{References}
\bibliographystyle{ieeetr}

\section{Appendix}
\subsection{Example of Determining the $\lambda_i$'s for a Single Linear Constraint}
\label{example1}
Here the procedure for a single constraint in addition to the necessary unit-trace constraint is illustrated. To fix ideas, perturbations at constant energy are assumed. Then, the constraining operators are  $\hat{C}_1 = \hat{G}_1 = \hat{I}$ and $\hat{C}_2 = \hat{G}_2 = \hat{H}$.  Eq. (\ref{Eq::Chap_4_gam_r_gam_eps}) simplifies to
\begin{equation}
	\hat{\gamma}_r = (1- 2\lambda_1)\,\hat{\gamma}_\epsilon- \{\hat{H},\hat{\gamma}_\epsilon\}\, \lambda_2 
\end{equation}
and the system of Eqs.                                                                                                                                                                                                                                                                                                                                                                                                                                                                                                                                                                                                                                                                                                                                                                                                                                                                                                                                                                                                                                                                                                                                                                                                                                                                                                                                                                                                                                                                                                                                                                                                                                                                                                                                                                                                                                                                                                                                                                                                                                                                                                                                                                                                                                                                                                                                                                                                                                                                                                                                                                                                                                                                                                                                                                                                                                                                                                                                                                                                                                                                                                                                                                                                                                                                                                                                                                                                                                                                                                                                                                                                                                                                                                                                                                                                                                                                                                                                                                                                                                                                                                                                                                                                                                                                                                                                                                                                                                                                                                                                                                                                                                                                                                                                                                                                                                                                                                                                                                                                                                                                                                                                                                                                                                                                                                                                                                                                                                                                                                                                                                                                                                                                                                                                                                                                                                                                                                                                                                                                                                                                                                                                                                                                                                                                                                                                                                                                                                                                                                                                                                                                                                                                                                                                                                                                                                                                                                                                                                                                                                                                                                                                                                                                                                                                                                                                                                                                                                                                                                                                                                                                                                                                                                                                                                                                                                                                                                                                                                                                                                                                                                                                                                                                                                                                                                                                                                                                                                                                                                                                                                                                                                                                                                                                                                                                                                                                                                                                                                                                                                                                            (\ref{Eq::Chap_4_gam_r_gam_eps_cons}) that determine the values of $\lambda_1$ and $\lambda_2$ becomes
\begin{equation}
\begin{aligned}
	(1-2\lambda_1)^2 \,g_1 +4(1-2\lambda_1)\,\lambda_2\,g_2+\lambda_2^2 \,g_3&= 1\\
		(1-2\lambda_1)^2\, g_2+ (1-2\lambda_1)\,\lambda_2\,g_3+\lambda_2^2 \,g_4&= E_0
	\label{Eq::Chap_4_Examp_Constr_1}
\end{aligned}
\end{equation}
where $E_0=\mathrm{Tr}(\hat{\rho}_0\hat{H})$ is the energy of the base state and for shorthand $g_1=\mathrm{Tr}(\hat{\gamma}_\epsilon^2)$, $g_2=\mathrm{Tr}(\hat{\gamma}_\epsilon^2H)$, $g_3=\mathrm{Tr}(\{\hat{H},\hat{\gamma}_\epsilon\}^2)$, and $g_4=\mathrm{Tr}(\{\hat{H},\hat{\gamma}_\epsilon\}^2H)$ are defined.

\subsection{Explicit Solution for the Case of No Nontrivial Constraints}
\label{Constraintschapunittrace}

For no constraints except the necessary unit-trace condition the procedure can be solved explicitly in terms of the representation of Hermitian operators discussed in Section \ref{Sec::generalperturbation}. Operator $\hat{\gamma}_\epsilon$ [Eq. (\ref{gammaepsilon})] has coefficients
\begin{equation}
	\eta[\hat{\gamma}_\epsilon ]_{i,j}=\delta_{ij}\,\eta[\hat{\gamma}_0] _{i,i}+\eta_{i,j} 
\end{equation}
where the base state $\eta[\hat{\gamma_0}_\epsilon ]_{i,i}$'s are given in Eq. (\ref{eta0}) and the $ \eta_{i,j}$'s are independently sampled from $\mathcal{N}(0, \tilde\sigma)$. The trace of its square, denoted for shorthand as $t_\epsilon^2$, is
\begin{equation}
	t_\epsilon^2=\mathrm{Tr}(\hat{\gamma}_\epsilon^2)= \sum_{i,j = 0}^3\left(\delta_{ij}\,\eta[\hat{\gamma}_0 ]_{i,i}+\eta_{i,j} \right)^2
\end{equation}

Eq. (\ref{Eq::Chap_4_gam_r_gam_eps}) simplifies to $	\hat{\gamma}_r = (1- 2\lambda_I)\,\hat{\gamma}_\epsilon$
and the system of Eqs.                                                                                                                                                                                                                                                                                                                                                                                                                                                                                                                                                                                                                                                                                                                                                                                                                                                                                                                                                                                                                                                                                                                                                                                                                                                                                                                                                                                                                                                                                                                                                                                                                                                                                                                                                                                                                                                                                                                                                                                                                                                                                                                                                                                                                                                                                                                                                                                                                                                                                                                                                                                                                                                                                                                                                                                                                                                                                                                                                                                                                                                                                                                                                                                                                                                                                                                                                                                                                                                                                                                                                                                                                                                                                                                                                                                                                                                                                                                                                                                                                                                                                                                                                                                                                                                                                                                                                                                                                                                                                                                                                                                                                                                                                                                                                                                                                                                                                                                                                                                                                                                                                                                                                                                                                                                                                                                                                                                                                                                                                                                                                                                                                                                                                                                                                                                                                                                                                                                                                                                                                                                                                                                                                                                                                                                                                                                                                                                                                                                                                                                                                                                                                                                                                                                                                                                                                                                                                                                                                                                                                                                                                                                                                                                                                                                                                                                                                                                                                                                                                                                                                                                                                                                                                                                                                                                                                                                                                                                                                                                                                                                                                                                                                                                                                                                                                                                                                                                                                                                                                                                                                                                                                                                                                                                                                                                                                                                                                                                                                                                                              (\ref{Eq::Chap_4_gam_r_gam_eps_cons}) that determine the value of $\lambda_I$  reduces to the unit-trace condition $ \mathrm{Tr}(\hat{\gamma}_r^2)= 1$, which yields $(1- 2\lambda_I)=1/t_\epsilon$, so that the coefficients of the non-negative square root $\hat{\gamma}_r$ of the desired perturbed density operator are

\begin{equation}\label{etagammarnoconstraints}
	\eta[\hat{\gamma}_r ]_{i,j}=\frac{\delta_{ij}\,\eta[\hat{\gamma}_0 ]_{i,i}+\eta_{i,j}}{t_\epsilon}   
\end{equation}

\subsection{Relation between the $\chi$ distribution and the $\eta_{i,j}$ values after C1 constraint}
\label{appendixA}
In this appendix, we discuss the relationship between the 16-dimensional set of random variables $\eta_{i,j}$, the theoretical $\chi$ probability distribution, and that based on the C1 constraint. 

\begin{figure}[h]
\includegraphics[width=9cm]{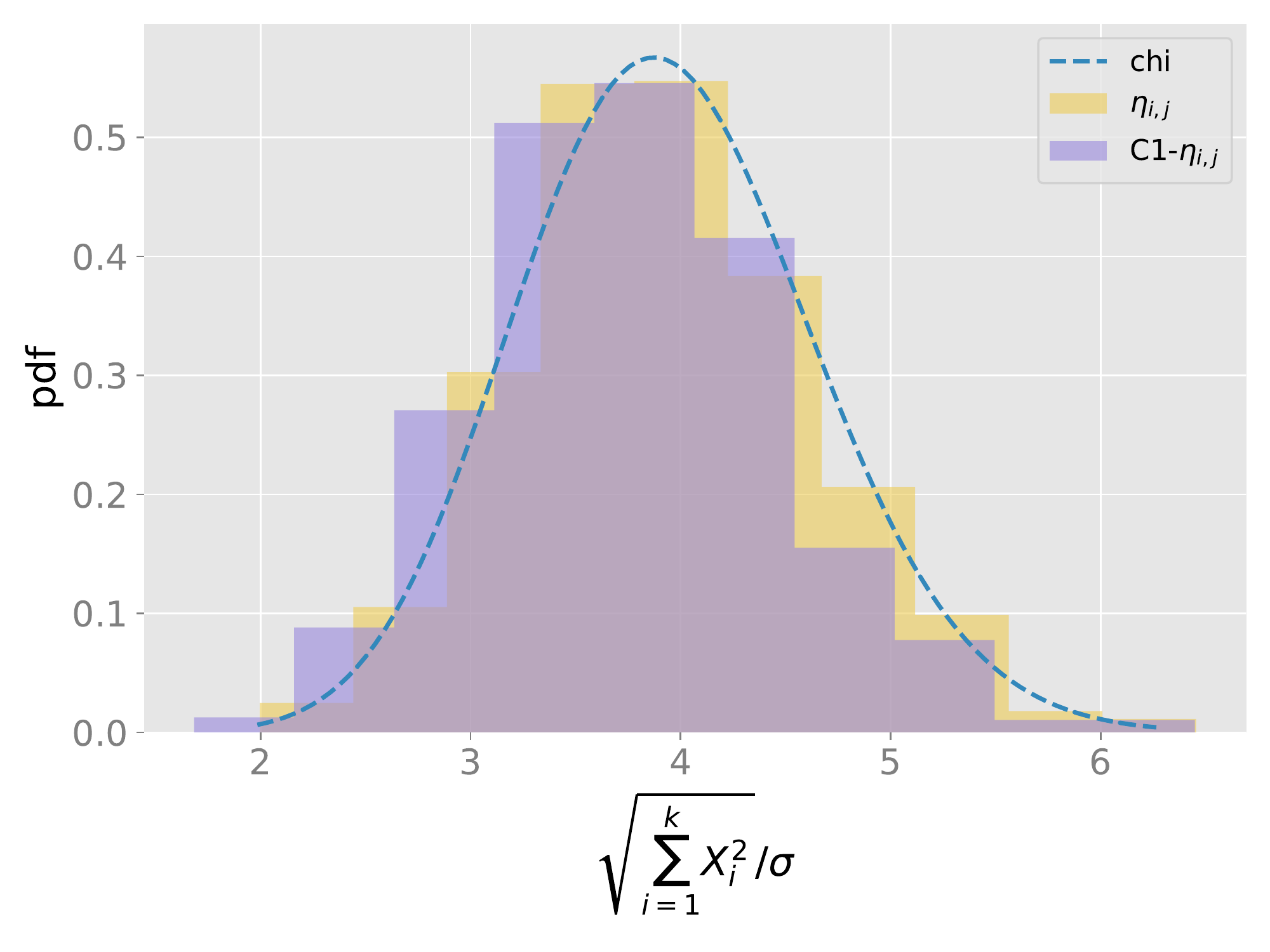}
\caption{\label{Figchi} Blue dotted curve: $\chi$ distribution ($k$=16). Yellow histogram: distribution of values of $\sqrt{X_i^2}/\sigma$ for 1,000 samples of a 16-dimensional set of random variables $X_i=\eta_{ij}$ all drawn from $\mathcal{N}(0, \sigma)$. Purple histogram:  distribution of the values $X_i=\eta[\hat{\gamma}_r ]_{i,j}$ obtained for the same samples via cyan Eq. (\ref{Eq18}) with the C1 constraint applied}. 
\end{figure}

To begin with, we recall that the positive square root of the sum of the squares of a set of independent random variables $\eta_{i,j}$ each picked from  $\mathcal{N}(0, 1)$ (i.e., from a standard normal distribution with zero mean, $ \mu_{i,j}=0$, and unit standard deviation, $\sigma_{i,j}=1$) is distributed according to the so-called $\chi$ distribution. When the $\eta_{i,j}$'s are taken from $\mathcal{N}(0, \sigma_{i,j})$, then $\eta_{i,j}/\sigma_{i,j}$ can be said to be equivalently taken from $\mathcal{N}(0, 1)$.   For example, if  all the $\sigma_{i,j}$ are equal, then 

\begin{equation}
	\left\langle
	\sqrt{\sum_{i,j = 0}^3 \eta_{i,j}^2} \right\rangle = \sqrt{2} \sigma\frac{\Gamma((k+1)/2)}{\Gamma(k/2)}
	\label{chi}
\end{equation}
where $k=16$ is the number of random variables. 
This is the case analyzed in Section \ref{Perturbation}. Fig. \ref{Figchi} presents the distribution (in yellow) for 1,000 random samples of the 16-dimensional set of variables $\eta_{i,j}$  all drawn from $\mathcal{N}(0, \sigma)$ with $\sigma=0.05$. The corresponding  $\chi$ distribution is indicated  by the dotted blue curve, with mean equal to 3.94.  

\begin{figure}[h]
\includegraphics[width=9cm]{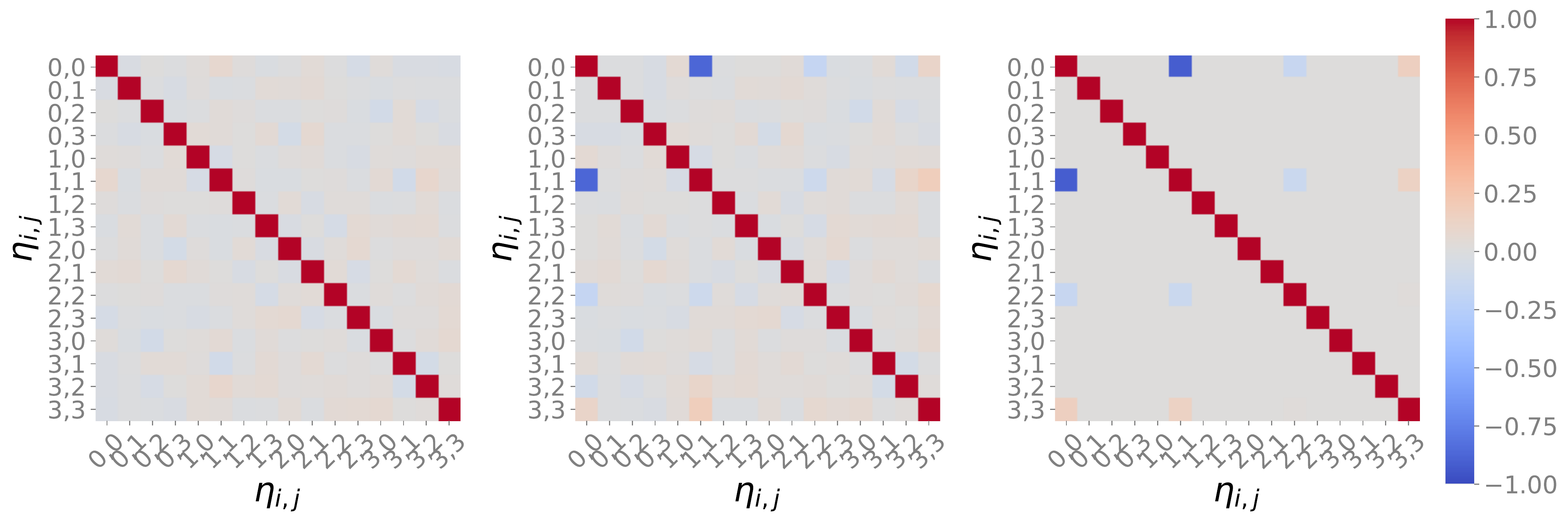}
\caption{\label{Figetasapp} Left: correlation diagram for 1,000 random samples of the $\eta_{i,j}$'s drawn from $\mathcal{N}(0, \sigma)$. Center: correlation diagram for the values obtained numerically with the C1 constraint. Right: correlation diagram for the transformed variables  $\eta[\hat{\gamma}_r ]_{i,j}$ obtained analytically from Eq. (\ref{correlationeq}) with the C1 constraint applied, i.e., the unit-trace condition $ \mathrm{Tr}(\hat{\gamma}_r^2)= 1$.}
\end{figure}

 Fig. \ref{Figchi} also presents, in purple, the distribution for the same samples  after the C1 constraint is applied. As can be seen, the distribution no longer corresponds to a $\chi$ distribution. This is because the C1 constraint induces strong correlations between $\eta_{0,0}$ and $\eta_{1,1}$ and between $\eta_{1,1}$ and $\eta_{0,0}$ as seen in Fig. \ref{Figetasapp}.
 
 These correlations can be also computed explicitly from the  solution obtained in Section \ref{Constraintschapunittrace}. In fact, Eq. (\ref{etagammarnoconstraints}) represents a nonlinear transformation from the  random variables $\eta_{i,j}$ to the perturbed state variables $\eta[\hat{\gamma}_r ]_{i,j}$.
 Since the $\eta_{i,j}$ are independent, their variance-covariance matrix $\Sigma$ is diagonal, with the variances $\Sigma_{i,j;i,j}=\sigma_{i,j}^2$ on the diagonal. But the covariance matrix of the transformed variables $\eta[\hat{\gamma}_r ]_{i,j}$ is given by $\mbox{Cov}=J\,\Sigma\,J^T$ where $J$ is the Jacobian of the transformation evaluated at the mean values $\mu_{i,j}$ of the original variables $\eta_{i,j}$, i.e., from Eq. (\ref{etagammarnoconstraints}),
 \begin{equation}\label{Jacobian}
 	J_{i,j;k,\ell}=\frac{\delta_{ik}\delta_{j\ell}}{t_\epsilon}-\frac{(\delta_{ij}\,\eta[\hat{\gamma}_0 ]_{i,i}+\mu_{i,j})( \delta_{k\ell}\,\eta[\hat{\gamma}_0 ]_{k,k} +\mu_{k,\ell})}{t_\epsilon^3} 
 \end{equation}
where $t_\epsilon= 1+2\sum_{i = 0}^3  \eta[\hat{\gamma}_0 ]_{i,i}\mu_{i,i} + 
\sum_{i,j = 0}^3 \mu_{i,j}^2$ and we recall that $\sum_{i = 0}^3 \eta[\hat{\gamma}_0 ]^2_{i,i}=1$.

When the 16 independent random variables $\eta_{i,j}$ are all drawn from $\mathcal{N}(0, 1)$, then $\mu_{i,j}=0$ and $\sigma_{i,j}=1$ for all $i$ and $j$, $\Sigma_{i,j;k,\ell}=\delta_{ik}\delta_{j\ell}$ and the covariance matrix simplifies to
 \begin{equation}
	\mbox{Cov}_{i,j;k,\ell}=\delta_{ik}\delta_{j\ell}-\,\delta_{ij}\,\eta[\hat{\gamma}_0 ]_{i,i}\, \delta_{k\ell}\,\eta[\hat{\gamma}_0 ]_{k,k}  
\end{equation}
and yields the correlation matrix
 \begin{equation}
	\mbox{Corr}_{i,j;k,\ell}=\frac{\delta_{ik}\delta_{j\ell}-\,\delta_{ij}\,\eta[\hat{\gamma}_0 ]_{i,i} \delta_{k\ell}\,\eta[\hat{\gamma}_0 ]_{k,k}  }{\sqrt{1-\,\delta_{ij}\,\eta[\hat{\gamma}_0 ]^2_{i,i}}\sqrt{1-\,\delta_{k\ell}\,\eta[\hat{\gamma}_0 ]^2_{k,k}}}\label{correlationeq}
\end{equation}

For our particular choice of the scalar coefficients, $c_0 = 1$, $c_1 = 0.996$, $c_2 = 0.4$, $c_3 = -0.4$, we have $a=0.0316$, $b=0.5468$, $c=0.8361$, $d=0.0316$, $\eta[\hat{\gamma}_0 ]_{0,0}=0.7231$, $\eta[\hat{\gamma}_0 ]_{1,1}=0.6598$, $\eta[\hat{\gamma}_0 ]_{2,2}=0.1446$, $\eta[\hat{\gamma}_0 ]_{3,3}=-0.1446$,  and the only nonzero off-diagonal entries of the (symmetric) correlation matrix are $\mbox{Corr}_{1,1;0,0}=-0.9191$, $\mbox{Corr}_{2,2;0,0}=-0.1530$,  $\mbox{Corr}_{3,3;0,0}=0.1530$, $\mbox{Corr}_{2,2;1,1}=-0.1283$,  $\mbox{Corr}_{3,3;1,1}=0.1283$, and $\mbox{Corr}_{3,3;2,2}=0.0214$.
Fig. \ref{Figetasapp} illustrates the above relations by comparing the correlation diagrams for a set of 1000 random samples of the $\eta_{i,j}$'s drawn from $\mathcal{N}(0, 1)$ and the corresponding values of the transformed variables  $\eta[\hat{\gamma}_r ]_{i,j}$.

\end{document}